# Risk and Scientific Reputation: Lessons from Cold Fusion

Huw Price[1]

University of Bonn and Trinity College, Cambridge

**Abstract:** Many scientists have expressed concerns about potential catastrophic risks associated with new technologies. But expressing concern is one thing, identifying serious candidates another. Such risks are likely to be novel, rare, and difficult to study; data will be scarce, making speculation necessary. Scientists who raise such concerns may face disapproval not only as doomsayers, but also for their unconventional views. Yet the costs of false negatives in these cases – of wrongly dismissing warnings about catastrophic risks – are by definition very high. For these reasons, aspects of the methodology and culture of science, such as its attitude to epistemic risk and to unconventional views, are relevant to the challenges of managing extreme technological risks. In this piece I discuss these issues with reference to a real-world example that shares many of the same features, that of so-called 'cold fusion'.

## 1 CSER and Maverick Science

Plans for the Cambridge Centre for the Study of Existential Risk (CSER) first emerged in conversations between Martin Rees, Jaan Tallinn and myself, with others, in Spring 2012. In one of those discussions I remarked to Rees that some of the issues we wanted CSER to study had a poor reputation. They were regarded as 'a bit flakey', as I put it. Rees agreed, but said that that was why the project was important. Serious risks might not be getting the attention they deserved, because of these reputational factors.

From that point, we were clear that a useful role for CSER might be to act as a reputational counterweight. In other words, CSER could use the reputation that we ourselves had at hand – that of Cambridge, and of our distinguished supporters and collaborators – as an opposing force, to nudge these neglected issues away from the fringes, towards respectability. In this way we could help to determine which neglected risks really needed attention and which could safely be left on the sidelines.

Three years later, when CSER won funding for the project *Managing Extreme Technological Risk* from the Templeton World Charity Foundation (TWCF)*,* the role of reputation in science was an explicit focus of one of five subprojects. This is how we presented this work in our application to TWCF:

---

[1] **Author bio:** Huw Price is Emeritus Bertrand Russell Professor of Philosophy and an Emeritus Fellow of Trinity College, Cambridge. He was co-founder with Martin Rees and Jaan Tallinn of the Cambridge Centre for the Study of Existential Risk, and Academic Director (2016–2021) of the Leverhulme Centre for the Future of Intelligence. From 2022 he is a Distinguished Professor Emeritus at the University of Bonn, Germany.

> *Extreme risk and the culture of science.* Prediction and mitigation of ETRs [extreme technological risks] is likely to depend on long-range evaluation of possibilities that seem far-fetched, in some cases. Many of these possibilities may turn out to be of negligible concern, but the net needs to be cast widely in the first instance, to maximise our chances of catching the fish that matter, as early as possible. Given the nature of the risks involved, there is a high cost to 'false negatives'.
>
> Unfortunately, science is not good at casting its net widely. As Kuhn (1962) observed, science is conservative, and there is strong cultural pressure on scientists to work within the current paradigm. Advances – Kuhn's 'scientific revolutions' – often depend on far-sighted individuals who resist these pressures, to work outside the mainstream. The history of science offers many examples of such figures, whose work is often shunned for long periods, before eventual vindication. Of course, history offers far more examples of fringe proposals that were not vindicated by later developments. In general, we rely on the normal process of science to sort out the gems from the dross – it may take a long time in some cases, but we get there in the end. In the special case of ETR, however, such a delay might be extremely costly. This subproject investigates this danger, and ways to reduce it.

Two years later again, in 2017, CSER organised a workshop on 'Risk and the Culture of Science'. It was held at Trinity College, Cambridge, in association with the When Experts Disagree (WEXD) Project based at University College, Dublin. We described the theme like this:

> Many scientists have expressed concern about potential catastrophic risks associated with powerful new technologies. But expressing concern is one thing, identifying serious candidates another. By definition, such risks will be novel, rare and difficult to study; data will be scarce, speculation necessary. This pushes us to the fringes of science, the realm of 'mavericks' and the unconventional – often a hostile and uncomfortable place. Scientists value consensus, at least about the big issues.
>
> Catastrophic risk is both a big issue and a highly charged one: so fringe-dwellers may be doubly unwelcome. Do we need to make special efforts to protect our mavericks, if catastrophic risk is to get the attention it deserves? If so, how can we do it? Can we use the values of science to protect useful fringe-dwellers from science's own immune system? Can we engineer a Maverick Room? (CSER 2017)

The workshop involved a number of leading philosophers of science, including Professor Heather Douglas (Michigan State), whose work on the intersection of epistemic risk and value in science had been one of the inspirations for the project (Douglas 2000, 2007, 2009) – more on Douglas's views below. It also included some speakers we called our mavericks – researchers who felt that they had encountered these reputational issues in their own work. They spoke about their own experience in controversial fields such as nanotechnology risk, AI risk, geoengineering, and so-called 'cold fusion'.

The last example was particularly interesting, from my personal point of view. At around the same time as plans for CSER were first emerging, I had happened to become interested in claims then being made about cold fusion, or LENR ('low energy nuclear reactions'), as it was also termed. By the time of the workshop in 2017 I had been following the field for several years. I regarded it as a fascinating real-world example of maverick science, in the sense we had in mind.

Like many millions of others, I had been aware of the claims of cold fusion after its public and controversial debut in 1989. I had kept an eye on it for some time afterwards, on the online forums provided in those days by internet newsgroups. But it had dropped off my radar for many years. Then, late in 2011, a new generation online forum – Facebook – happened to make it visible to me again. As I'm going to explain, I have followed it ever since, writing several public pieces about it, and meeting a number of the leading figures in the field, both inside and outside academia.

Nothing in those ten years has shaken my conviction that cold fusion is a fascinating real-world example of maverick science, in the sense relevant to the study of extreme technological risk. Indeed, I have come to see my own experience in thinking and writing about the field – in particular, some of the reactions I have encountered from others – as an interesting illustration of some of the general characteristics of maverick science. The present piece is a kind of ten year progress report, from my personal and professional point of view – that of a philosopher of science, with an interest in the science of extreme risk. I'll tell the story of my own engagement with the field, and describe the lessons I think that we should take from it.

I'll tell this story, in part, by reproducing three of my public pieces, from 2015, 2017 and 2019 – they appear below as the starred sections §2*, §4* and §6*; readers familiar with these pieces may skip over them, of course. (In the present version of the paper these sections are also shaded.) I'll fill out the narrative provided by these pieces with some additional detail about my engagement with the field at various points, and about developments in the field itself, especially in the years since 2019. I'll close with an assessment of where I think the field stands, and what lessons I think the case carries for risk and the culture of science.

My first public piece appeared in the online magazine *Aeon* in December 2015 (Price 2015). The text below is from a lightly edited version I prepared for a WEXD workshop in Dublin, in July 2017. Apart from three new introductory paragraphs – the first of them sadly apt, in the light of recent events – the main difference is that the *Aeon* version left out some of my original references and the Dublin version put these back in. I have tweaked the tenses in a few places, and added citations formerly provided by hyperlinks. And I have restored the title that I gave the piece originally in 2015 – *Aeon* preferred something a little less obscure! The Dublin version also included a postscript from 2017, with some updates about the field. It appears separately below (§4*). Apart from that, I have resisted the temptation to update the piece with the benefit of hindsight.

## 2* My Dinner with Andrea (December 2015)

Science can be unkind to its mavericks – often they are shunned, even ridiculed. Sometimes, like the Australians Barry Marshall and Robin Warren (who eventually won a Nobel prize for their discovery that stomach ulcers are caused by a bacterium), the mavericks get the last laugh. But should they have had to endure all those first laughs? We would have had better ulcer treatments sooner if Marshall and Warren had been listened to earlier, presumably. It is easy to imagine how the costs of delay might be much more severe. Suppose a maverick who had identified some previously unnoticed pathway to a new pandemic – millions of lives might hang on the question of who was laughing at whom.

I'm a philosopher of science, and also co-founder of Cambridge's Centre for the Study of Existential Risk, so pessimistic speculations like these are part of my job. But in recent years I've been watching a real-life case. I've been following the fortunes of one of the most widely-ridiculed groups in science (or pseudoscience, as their critics would say), the advocates of so-called 'cold fusion'. This community lives in what I now call a reputation trap, tainted with a reputation so dismissive, and so contagious, that the entire area is effectively off-limits to mainstream investigators. Anyone who is seen to approach the area with an open mind risks being pushed into the trap themselves – I speak from experience!

I'm increasingly convinced that the culture of reputation traps is unhealthy for science, and potentially dangerous for all of us, at least in cases where the costs of 'wrongful dismissal' are especially high. In particular, I'm worried that science itself is caught in a kind of meta trap. It is unable to consider the possibility that it is insufficiently open-minded, because the reputational fences themselves prevent it from asking the question whether the fences are too high, or erected in the wrong places.

My own journey towards the fringes began in 2011, when a physicist friend made a joke on Facebook about the number of laws of physics being broken in Italy. He had two pieces of news in mind. One was a claim by the Gran Sasso-based OPERA team to have discovered super luminal neutrinos (Reich 2011). The other concerned an engineer from Bologna called Andrea Rossi, who claimed to have a cold fusion reactor producing commercially useful amounts of heat (Hambling 2011).

Why were these claims so improbable? The neutrinos challenged a fundamental principle of Einstein's theory of special relativity, that nothing can travel faster than light. While cold fusion, or LENR, is the controversial idea popularised by Martin Fleischmann and Stanley Pons in 1989, that nuclear reactions similar to those in the sun could also occur at or close to room temperature, under certain conditions. Fleischmann and Pons claimed to have found evidence that such reactions could occur in palladium loaded with deuterium (an isotope of hydrogen). A few other physicists, including Sergio Focardi at Bologna (Wikipedia contributors, 2021a), claimed similar effects with nickel and ordinary hydrogen. But most were highly sceptical, and the field 'subsequently gained a reputation as pathological science,' as Wikipedia puts it (Wikipedia contributors, 2021b). Even the believers had not claimed commercially useful quantities of excess heat, as Rossi now reported from his 'E-Cat' reactor.

It turned out that my physicist friend and I disagreed about which of these unlikely claims was the less improbable. He thought the neutrinos, on the grounds that the work had been done by respectable scientists, rather than by a lone engineer with a somewhat chequered past. I thought Rossi, on grounds of the physics. Superluminal neutrinos would overturn a fundamental tenet of relativity, but all Rossi needed was a previously unnoticed channel to a reservoir of energy whose existence is not in doubt. We know that huge amounts of energy are locked up in metastable nuclear configurations, trapped like water behind a dam. There's no known way to get useful access to that energy, at low temperatures. But – so far as I knew – there was no 'watertight' argument that no such method exists.

My friend agreed with me about the physics. (So has every other physicist I've asked about it since.) But he still put more weight on the sociological factors – reputation, as it were. So we agreed to bet a dinner on the issue. My friend would pay if Rossi turned out to have something genuine, and I would pay if the neutrinos came up trumps. We'd split the bill if, as then seemed highly likely, both claims turned out to be false.

It soon became clear that I wasn't going to lose. The neutrinos were scratched from the race, when it turned out that someone on OPERA's team of respectable scientists had failed to tighten an optical lead correctly (Reich 2012). Rossi, however, seemed to be going from strength to strength.

While it is fair to say that the jury is still out (literally, as we'll see), there has been a lot of good news, for my hopes of a free dinner, in the past few years. There have been two reports (Levi et al. 2013, 2014) of tests of Rossi's device by teams of Swedish and Italian physicists whose scientific credentials are not in doubt, and who had access to one of his devices for extended periods (a month, for the second test). Both reports claimed levels of excess heat far beyond anything explicable in chemical terms, in the testers' view. (The second report also claimed isotopic shifts in the composition of the fuel.) Since then there have been several reports of duplications by experimenters in Russia (Lewan 2015a) and China (Wang 2015), guided by details in the 2014 report.

More recently (by late 2015), Rossi was granted a US patent for one of his devices, previously refused on the grounds that insufficient evidence had been provided that the technique worked as claimed (Lewan 2015b). There were credible reports that a 1MW version of his device, producing many times the energy that it consumes, had been on trial in an industrial plant in Florida for months, with good results so far. And Rossi's US backer and licensee, Tom Darden – a respectable North Carolina-based industrialist, with a long track record of investment in pollution-reducing industries – has been increasingly willing to speak out in support of the LENR technology field (Dumaine 2015).

Finally, there was a paper by two senior Swedish physicists, Rickard Lundin and Hans Lidgren, proposing a mechanism for Rossi's results, inspired in part by the second of two test reports mentioned above (Lundin & Lidgren 2015). Lundin and Lidgren say that the 'experimental results by Rossi and co-workers and their E-Cat reactor provide the best experimental verification' of the process they propose.

As I say, I don't claim that this evidence is conclusive, even collectively. It's still conceivable that there is fraud involved, as many sceptics have claimed; or some large and persistent measurement error. Yet as David Bailey and Jonathan Borwein pointed out (Bailey and Borwein 2014, 2015), these alternatives seemed to be becoming increasingly unlikely – which seemed great news for my dinner prospects.

Moreover, Rossi was not the only person claiming commercially relevant results from LENR. Another prominent example is Robert Godes, of Brillouin Energy, profiled in a recent Norwegian newspaper piece (Bjørkeng 2015). If you want to dismiss Rossi on the grounds that he's claiming something impossible, one of these explanations (i.e., fraud or large and persistent error) needs to work for Godes, too.

You can see why I've been salivating at the thought of My Dinner with Andrea, as I've been calling it (h/t Louis Malle 1981), in honour of the man who would be the absent guest of honour, if my physicist friend is paying. And it is not only my stomach that has been becoming increasingly engaged with this fascinating story. I'm a philosopher of science, and my brain has been finding it engrossing, too.

What do I mean? Well, it hasn't escaped my attention that there's a lot more than a free dinner at stake. Imagine that someone had a working hot fusion reactor in Florida – assembled, as Rossi's 1MW device is reported to be, in a couple of shipping containers, and producing several hundred kilowatts of excess power, month after month, in apparent safety. That would be huge news, obviously. (As several people have noticed, a new clean source of energy would be really, really useful, right about now!)

But if the potential news is this big, why haven't most of you heard about Rossi, or Godes, or any of the other people who have been working in the area (for many years, in some cases)? This is where things get interesting, from a philosopher of science's point of view.

As a question about sociology, the answer is obvious. Cold fusion is dismissed as pseudoscience, the kind of thing that respectable scientists and science journalists simply don't talk about (unless to remind us of its disgrace). As a recent *Fortune* piece puts it, the Fleischmann and Pons 'experiment was eventually debunked and since then the term cold fusion has become almost synonymous with scientific chicanery.' (Dumaine 2015). In this case, the author of the article is blithely reproducing the orthodox view, even in the lead-in to his interview with Tom Darden – who tells him a completely different story (and has certainly put his money where his mouth is).

Ever since 1989, in fact, the whole subject has been largely off-limits, in mainstream scientific circles and the scientific media. Authors who do put their head above the parapet are ignored or rebuked. Most recently, Lunden and Lidgren report (Lewan 2015c) that they submitted their paper to the journal *Plasma Physics and Controlled Fusion,* but that the editors declined to have it reviewed; and that even the non-reviewed preprint archive, arxiv.org, refused to accept it.

So, as a matter of sociology, it is easy to see why Rossi gets little serious attention; why an interview with Tom Darden associates him with scientific chicanery; and why, I hope, some of you are having doubts about me, for writing about the subject in a way that indicates that I am prepared to consider it seriously. (If so, hold that attitude. I want to explain why I take it to reflect a pathology in our present version of the scientific method. My task will be easier if you are still suffering from the symptoms.)

Sociology is one thing, but rational explanation another. It is very hard to extract from this history any satisfactory *justification* for ignoring recent work on LENR. After all, the standard line is that the rejection of cold fusion in 1989 turned on the failure to replicate the claims of Fleischmann and Pons. Yet if that were the real reason, then the rejection would have to be provisional. Failure to replicate couldn't possibly be more than provisional – empirical science is a fallible business, as any good scientist would acknowledge. In that case, well-done results claiming to overturn the failure to replicate would certainly be of great interest.

Perhaps the failure to replicate wasn't crucial after all? Perhaps we knew on theoretical grounds alone that cold fusion was impossible? But this would make nonsense of the fuss made at the time and since, about the failure to reproduce the Fleischmann and Pons results. And in any case, it is simply not true. As I said at the beginning, what physicists actually say (in my experience) is that although LENR is highly unlikely, we cannot say that it is impossible. We know that the energy is in there, after all.

No doubt one could find some physicists who would claim it was impossible. But they might like to recall the case of Lord Rutherford, greatest nuclear physicist of his day, who famously claimed that 'anyone who expects a source of power from transformation of . . . atoms is talking moonshine' – the very day before Leo Szilard, prompted by newspaper reports of Rutherford's remarks, figured out the principles of the chain reaction that makes nuclear fission useable as an energy source, peaceful or otherwise. (The story is told in Rhodes 1986, ch. 1.)

This is not to deny that there is truth in the principle popularised by Carl Sagan, that extraordinary claims require extraordinary evidence. We should certainly be very cautious about such surprising claims, unless and until we amass a great deal of evidence. But this is not a good reason for ignoring such evidence in the first place, or refusing to contemplate the possibility that it might exist. (As Robert Godes said recently, 'It is sad that such people say that science should be driven by data and results, but at the same time refuse to look at the actual results'; Bjørkeng 2015.)

Again, there's a sociological explanation why few people are willing to look at the evidence. They put their reputations at risk by doing so. Cold fusion is tainted, and the taint is contagious – anyone seen to take it seriously risks contamination themselves. So the subject is stuck in a place that is largely inaccessible to reason – a reputation trap, we might call it. People outside the trap won't go near it, for fear of falling in. 'If there is something scientists fear it is to become like pariahs,' as Rickard Lundin puts it (Lewan 2015c). People inside the trap are already regarded as disreputable, an attitude that trumps any efforts they might make to argue their way out, by reason and evidence.

Outsiders might be surprised how well-populated the trap actually is, in the case of cold fusion and LENR. The field never entirely went away, nor vanished from the laboratories of respected institutions. Rossi's own background is not in these laboratories, but he acknowledges that his methods owe much to those who are, or were – especially to the late Sergio Focardi, one of the pioneers of the field (Wikipedia contributors 2021a). To anyone willing to listen, the community will say that they have amassed a great deal of evidence of excess heat, not explicable in chemical terms, and of various markers of nuclear processes. Some, including a team at one of Italy's leading research centres, say that they have many replications of the Fleischmann and Pons results (ENEA 2013).[2]

Again, the explanation for ignoring these claims cannot be that other attempts failed, twenty five years ago. That makes no sense at all. Rather, it's the reputation trap. The results are ignored because they concern cold fusion, which we 'know' to be pseudoscience – we know it because attempts to replicate these experiments failed, twenty five years ago! The reasoning is still entirely circular, but the reputation trap gives its conclusion a convincing mask of respectability. That's how the trap works.

Fifty years ago, Thomas Kuhn (1962) taught us that this is the usual way for science to deal with paradigm-threatening anomalies. The borders of dominant paradigms are often protected by reputation traps, which deter all but the most reckless or brilliant critics. If LENR were an ordinary piece of science (or proposed science), the challenge by Rossi and others would provide some fascinating spectator sport for philosophers and historians of science – a Kuhnian revolution waiting to happen, perhaps, with threats to the stability of the reputation trap now clearly in view. We could take our seats on the sidelines, and wait to see whether walls fall –

---

[2] I was delighted to discover recently that some of the early Italian cold fusion research took place in the very same underground laboratory as the 2011 work that (briefly) reported superluminal neutrinos. The story is told in a fascinating 1994 essay by the Caltech physicist David Goodstein, who found himself in the position of having professional colleagues in both camps, in the heated early controversy about cold fusion. Here he describes some of the work of the ENEA team led by his friend Professor Francesco Scaramuzzi:

> Reacting to criticism of the primitive technique they had used to detect neutrons, they purchased the best neutron detection system in the world, essentially identical to the one used by Charlie Barnes at Caltech. Going one better, they installed it in physics laboratories that had been excavated under a mountain called the Gran Sasso, a two-hour drive from Rome. Anywhere on the surface of the Earth, there are always some neutrons buzzing around due to cosmic radiation from outer space. This so-called "background" has to be subtracted from the neutrons produced by any other phenomenon such as Cold Fusion. In the galleries under the Gran Sasso, the shielding effect of the mountain reduces the cosmic ray neutron background nearly to zero. That's why the laboratory was built there. An automated system was set up to monitor the neutron counter while running the temperature of a Scaramuzzi-type deuterium gas cell up and down. Every week or so, a member of the group would have to drive out to the Gran Sasso lab, check out the counters, replenish the supply of liquid nitrogen, and bring back the data. No one could accuse them any longer of being unsophisticated about neutron work. However, this experiment, like their own earlier work and many others blossoming around the world, produced positive results, but only sporadically. There was no dependable recipe for coaxing bursts of neutrons out of the Cold Fusion cell. (Goodstein 1994) whether distinguished sceptics end up with egg on their faces. 'Pleasant is it to behold great encounters of warfare arrayed over the plains, with no part of yours in the peril', as Lucretius put it (Lucretius 1916).

This would be plenty to explain why I've been finding Rossi's apparent progress so engaging. But there's more, much more. None of us, even philosophers, are mere spectators in this case. We all have skin in the game, and parts, indeed a planet, quite seriously in the peril. We are like a thirsty town, desperate for a new water supply. What we drink now is slowly killing us. We know that there's an abundant supply of clean, cheap water, trapped behind the dam. The problem is to find a way to tap it. A couple of engineers thought that they had found a way, twenty five years ago, but they couldn't make it work reliably, and the profession turned against them. Since then, there's been a big reputation cost to any engineer who takes up the issue.

Put this way, it is easy to see an argument that we've been shooting ourselves in the foot. In a case like this there is very little cost to a false positive – to investing some time and money in an avenue that turns out eventually to go nowhere. But there may be a huge cost to a false negative. If Rossi, Godes, Lunden, Lidgren and others do turn out to have something useful – something that can make some useful contribution to meeting our desperate need for clean, cheap energy – we will have wasted a generation of progress, by dumping cold fusion in the reputation trap. What we should have done instead is to have engineered the exact opposite of a reputation trap – an X Prize-like reward for the first reliable replication of the Fleischmann and Pons results, above some commercial bar, perhaps.

Now I can explain what I meant earlier, when I asked you to hold on to the thought that I must be a bit flakey myself, if that's your reaction to my willingness to take cold fusion seriously. If you do think that – at least if you think it without having studied the evidence for yourself – then your reaction is a symptom of the reputation trap. But I've now suggested that the trap itself may be an irrational pathology, in a special case like this, where the cost of a false negative is very high. If I'm right, then in a more rational world, we would fix our scientific norms to escape it. In a more rational world, you wouldn't think I'm flakey.

I don't need to deny that your reaction is an appropriate one, by the standards of science as we currently practice it, or that those standards work pretty well, in general. Reputation traps have a useful purpose, in the Kuhnian picture. They help to maintain the stability important to what Kuhn called normal science – the ordinary, useful kind of science when paradigms are not under threat. But this is compatible with the claim that they can be harmful in special cases (of which cold fusion may be one) – and that we could do better, if we were better at identifying those cases in advance.

I suspect it's too late to dismantle the trap for LENR – the horse is already in the process of bolting, I think. If so, then the field is going to be mainstream soon, in any case. But we could try to learn from our mistakes. There may be other potential cases with a similar payoff structure (a high cost for false negatives, with a low cost for false positives). I suspect there are some in the area of emerging extreme risks, another field in which I have some interest. I've met scientists employed as technology forecasters for a very large national enterprise, who were frustrated at their inability to persuade their organisation to list LENR as a technology of possible strategic interest, even with low probability. Once again, it was the reputation trap at work. This attitude could be disastrous in other cases, if the ideas stuck in the trap are the key to early detection of some potentially catastrophic risk. (Indeed, as my physicist friend rightly points out, it might be disastrous in the case of LENR itself, if it turns out to generate its own extreme risks.)

There are big issues here. Would it be possible to avoid these counterproductive cases of the reputation trap, without erring too far in the other direction – without opening the floodgates, so to speak? I'm not sure, but I think it is important to put the question on the table. If LENR does develop in the direction I now think likely, we might be able to salvage something useful from mistakes in this case.

I'll close with some words from Tom Darden, from a speech at ICCF-19, the international meeting of the LENR community held in Padua, Italy in April 2015. (A full transcript of the speech is accessible at McDonough 2015.) If the field is indeed in the process of digging itself out of the reputation trap, then Darden deserves much of the credit. Either way, his attitude displays the kind of cautious open-mindedness that has been so lacking in reactions to the field for most of its history.

Darden describes how he came to invest in LENR. Until quite recently, he says, he shared the conventional view that 'the subject was dead'. But several independent enquiries about LENR within a matter of weeks convinced him that there was something worth investigating:

> We believed LENR technology was worth pursuing, even if we turn out to be unsuccessful ultimately. We were willing to invest time and resources to see if this might be an area of useful research in our quest to eliminate pollution. At the time, we were not especially optimistic, but the global benefits were compelling.

He reports that things have been going very well. 'We've had some success, and we're expanding our work … and believe that we may be, at last, on the verge of a new paradigm shift.' Finally, he expresses his appreciation to his audience, the LENR community themselves, and recalls another controversial piece of Italian science:

> I would like to say how truly sorry I am that society has attacked you for the last three decades. The treatment of Fleischmann and Pons, and the treatment of many of you, by mainstream institutions and the media will go down in history as one more example of scientific infanticide, where entrenched interests kill off their divergent progeny. This seems to be a dark component of human nature, and I note the irony of it – we are in Padova, Galileo's home.

It would be easy to overstate the analogy between mainstream institutions and the Inquisition, but it isn't entirely empty. If we refuse to acknowledge the possibility that existing scientific institutions are not working as well as they might, we do something to reinforce it. If the reputation trap makes it impossible to question the role of the reputation trap, then the cardinals are winning.

**3 The View from 2016**

The above piece appeared in November 2015. I published a short update in *Aeon* in March 2016 (Price 2016), when I was even more struck by the lack of attention that the Rossi saga was getting, as the end of the claimed one-year test of his reactor approached. In my view, it was a fascinating story whatever the outcome. Even if it was fraud, it was an extraordinarily elaborate and long-lasting fraud, in which the victims included some well-known individuals and financial institutions. Yet there was virtually no coverage of the story, even from this angle. I took this to be a further manifestation of the reputation trap. One couldn't write about the story without putting on the table the view obviously held by Tom Darden and other financial backers, that LENR might not be pseudoscience after all – and that view was taboo, by the standards of the reputation trap.

More in a moment about what became of this part of the story. As I noted earlier, I updated my original *Aeon* article in May 2017. At that point I had two new audiences in mind. The first, as I mentioned, was a workshop organised by the When Experts Disagree (WEXD) Project at University College, Dublin, in July 2017 (McKay 2017); WEXD had been partners in our own 'mavericks' workshop at Trinity College, Cambridge, in April 2017. I'll come back to the second audience later. The new version included the following postscript, which begins with a brief update about Rossi's case.

**4\* Postscript (May 2017)**

It's now more than a year since I wrote about these ideas in *Aeon*. How has LENR been faring?

About Andrea Rossi's e-Cat, I said above that the jury was still out. That's now become literally true! The rumoured one-year test of Rossi's 1MW reactor was completed in Spring 2016. As predicted, the verdict of the appointed 'expert responsible for validation' (ERV) was strongly positive. The ERV reported heat outputs many times the input energy, throughout the course of the test. However, Rossi's financial backers, Tom Darden's company Industrial Heat (IH), were not convinced. They declined to pay the $89m success fee to which Rossi claims they were contractually committed. Rossi took them to court, and that is where things stand.

The case comes before a jury in Florida in June 2017, and the pre-trial discovery process has already revealed a fascinating mass of information about Rossi's plant, about his history with IH, and about their interest in LENR in general. IH want to distance themselves from Rossi, but show no sign of regretting their commitment to the field as a whole. On the contrary, they appear to be seeking to accumulate as much as possible of the potential IP – as any rational investor would, presumably, if convinced that there was something there.

Elsewhere in the LENR field there have been several interesting new announcements of positive results. One of the most striking is a report by the respected lab SRI International, based in Menlo Park, California, of reactors provided by Robert Godes' Brillouin Energy. Brillouin

Energy was the second of the two claimed LENR developers I mentioned in my original article. The SRI tester, Dr Francis Tanzella, summarizes his results as follows:

> We report here on the most recent nine months of extensive testing in Brillouin's two original [reactors] operated at its Berkeley laboratory, and in the past two months, with the second unit having been further situated at SRI. Brillouin has manufactured five identical metallic cores and has consecutively tested each one of them in its two [reactors], seemingly producing the same controlled heat outputs repeatedly.
>
> Since its reconstruction and calibration, I have been able to corroborate that the [reactor] system moved to SRI continues to produce similar LENR Reaction Heat that it produced up in its Berkeley laboratory at Brillouin. Together with my prior data review, it is now clear that these very similar results are independent of the system's location (Berkeley or Menlo Park) or operator (Brillouin's or SRI's personnel). This transportable and reproducible reactor system is extremely important and extremely rare. These two characteristics, coupled with the ability to start and stop the reaction at will are, to my knowledge, unique in the LENR field to date. (Tanzella 2016, 2)

But this SRI report is not the only recent claim of reliable and reproducible results. Another came from the Condensed Matter Nuclear Science (CMNS) Department at Tohoku University, Japan, a program supported by the Japanese Clean Planet initiative. They, too, report 'highly reproducible' excess heat (Kaneko 2016).

The Clean Planet initiative was established in 2012, after the Fukushima disaster the previous year, and is an enthusiastic backer of research on LENR. The reputation trap seems always to have been shallower in Japan, where there is a long history of LENR research by groups including Mitsubishi Heavy Industry Ltd., and what there was of it seems to have been swept away by the tragic events of 2011.[3]

In the West it is still a different matter, but there are a few new cracks in the wall. In September 2016 *New Scientist* published an article entitled 'Cold fusion: Science's most controversial technology is back', reporting in a surge of commercial interest in LENR:

> You won't hear the words "cold fusion", but substantial sums of money are quietly pouring into a field now known as low-energy nuclear reactions, or LENRs. Earlier this year, the US House of Representatives Committee on Armed Services declared it was "aware of recent positive developments" in developing LENRs and noted their potential to "produce ultra-clean, low-cost renewable energy"'and their "strong national security implications". Highlighting too the interest of Russia, China, Israel and India, it

---

[3] Professor Jirohta Kasagi (Tohoku University) has recently informed me that in his view, the reputation trap has not been appreciably shallower in Japan. He surveys the early history of Japanese cold fusion research in (Kasagi and Iwamura 2008).

suggested the US could not afford to be left behind, and requested that the Secretary of Defense provide a briefing. (Brooks, 2016)

As if to cover its reputational bases, *New Scientist* accompanied this piece with an unfriendly editorial, dismissing the suggestion that there is now a case for public funding of LENR research.

> Taxpayer money could provide credibility and ensure that results are properly reported, rather than just rumoured. And if there is anything to cold fusion, it would be in the public interest for it to be investigated properly. But that's an enormous if. There's still no compelling reason to think cold fusion will work. Let those with money to burn take the risk and, if proven right, the rewards for their chutzpah too. For the rest of us, cold fusion is better off left out in the cold. (Anonymous, 2016)

In my view, this editorial is a sad example of the irrational pathology of the reputation trap. Why? Well, imagine an analogy (h/t Ridley Scott 2015). You are marooned on Mars when the rest of the expedition leaves without you. Food is running short, but you could improve your chances by growing your own. Unfortunately the only available source of fertiliser is the packaged waste from your former teammates – the crap they gave you before they left, as it were. There's 'no compelling reason' to think that this malodorous option will actually save you, but it's the best option in a tight corner. You hold your nose and get to work.

Let's agree with *New Scientist* that there's still no compelling reason to think cold fusion *will* solve our energy crisis. That's a very different matter from having a compelling reason to think that it *won't*. Our planet is in a very, very tight corner. How high does the probability need to be, before it makes sense to hold our reputational noses and join the search?

If there's one important difference between these two cases, it is that the stench of the reputation trap is an artefact of the culture of science itself, not an inevitable product of the biophysics of human by-products. If we agreed to take LENR seriously, the odour would simply vanish. A better analogy would be with the castaway who starves to death on a fertile island because hunting and gathering are taboo for a man of his high caste. As in that case, the obstacle to doing the rational thing is one we have created for ourselves.[4]

When *New Scientist* says that 'not many respectable scientists would touch cold fusion', it is playing its own role in the cultural process that maintains the obstacle, adding a new small smear to the opprobrium with which cold fusion is coated. The resulting odour keeps us all in a trap, not only by preventing us from exploring a significant part of a desperately-needed solution space, but also – at what philosophers call the meta level – by discouraging us even from raising the possibility that such a prohibition might be deeply irrational.

---

[4] In the light of my use of the Martian example, and my remark above that the reputation trap seemed to be shallower in Japan, I was delighted to learn recently that 'in eighteenth century Japan, biosolids were an esteemed substance'. Due to its importance as a fertiliser, 'Japanese citizens did not view human waste as unwanted muck, but rather as something of value' (Zeldovich 2019). So there is less difference between the Martian case and the high caste case than I imagined.

**5 Meeting Some of the Mavericks (2017–2018)**

As I said, I wrote the updated version of my original *Aeon* piece with two audiences in mind. One was the WEXD workshop at University College, Dublin, in July 2017. The second was proposed by the Australian science journalist, Mikey Slezak (a former student of mine at the University of Sydney). Slezak suggested that something based on my *Aeon* piece be included in the 2017 edition of an annual collection of Australian science writing. I gave him the updated version, with the postscript. In what I took to be yet another illustration of the reputation trap at work, it was then vetoed by a physicist on the committee, and never reached that audience.

A big benefit of the original *Aeon* piece, from my perspective, was that it brought me to the attention of many people within the LENR field. It also caught the eye of Clive Cookson, the *Financial Times* science editor who, as a young journalist, had written the very first newspaper story about the Fleischmann and Pons announcement in 1989. As Cookson (2012) recounts, he knew Martin Fleischmann at that point because his father had been a colleague of Fleischmann's in Chemistry at Southampton. Fleischmann sought his advice, as a journalist and family friend, before the famous news conference – which Cookson advised him against.

Within the field, one of my initial contacts was with Alan Smith, a very active and well-connected UK-based researcher. Smith was the cold fusion maverick at our workshop in Cambridge in 2017. He introduced me to others in the field, including some of the leading Japanese researchers, whom I visited in Sendai in October 2017. They showed me some of their experiments, including cells that they said were reliably producing a few watts of excess heat, for weeks at a time.

I also visited Stockholm, in November 2017, for what was billed as a demonstration of the latest version of Rossi's e-Cat. I met Andrea Rossi himself, and some of the Swedish and Italian scientists who had taken an interest in his work. Earlier in 2017, a few weeks after my postscript was written, Rossi had withdrawn his case against Industrial Heat, in an out of court settlement. Certainly, what had emerged in the legal process before that did not inspire great confidence that he had what he claimed. Nor, in my view, did the Stockholm demonstration.

Let's now pick up the story with my second public piece (Price 2019). This was written with an eye to the thirtieth anniversary of the infamous Fleischmann and Pons news conference, which fell in March 2019. By the time I wrote this piece I had been introduced by email to the SRI researcher mentioned above, Francis Tanzella, whose private comments reinforced the impression I had about Brillouin's progress from public material. I had also been introduced to one of the major figures of the LENR world, the former SRI team leader, Dr Michael McKubre; more on him below. Again, I have edited this piece as lightly as possible, mainly to add references.

**6\* Icebergs in the Room? Cold Fusion at 30 (March 2019)**

From aviation to zoo-keeping, there's a simple rule for safety in potentially hazardous pursuits. Always keep an eye on the ways that things could go badly wrong, even if they seem unlikely. The more disastrous a potential failure, the more improbable it needs to be before we can safely ignore it. Think icebergs and frozen O-rings (Wikipedia contributors, 2021c). History is full of examples of the costs of getting this wrong.

Sometimes the disaster is missing something good, not meeting something bad. For hungry sailors, missing a passing island can be just as deadly as hitting an iceberg. So the same principle of prudence applies. The more we need something, the more important it is to explore places we might find it, even if they seem improbable.

We desperately need some new alternatives to fossil fuels. To meet growing demands for energy, with some chance of avoiding catastrophic climate change, the world needs what Bill Gates called an energy miracle – a new carbon-free source of energy, from some unexpected direction. In this case it's obvious what the principle of prudence tells us. We should keep a sharp eye out, even in unlikely corners.

Yet there's one possibility that has been in plain sight for thirty years, but remains resolutely ignored by mainstream science. It is so-called *cold fusion,* or *LENR* (for Low Energy Nuclear Reactions). Cold fusion was made famous, or some would say infamous, by the work of Martin Fleischmann and Stanley Pons in March 1989. Fleischmann and Pons claimed to detect excess heat at levels far above anything attributable to chemical processes, in experiments involving the metal palladium, loaded with hydrogen. They concluded that it must be caused by a nuclear process – 'cold fusion', as they termed it.

Many laboratories failed to replicate Fleischmann and Pons' results, and the mainstream view since then has been that cold fusion was 'debunked'. It is often treated as a classic example of disreputable pseudoscience. But it never went away completely. It has always had defenders, including some scientists at very respectable laboratories. They acknowledged that replication and reproducibility were difficult in this field, but claimed that most attempts on which the initial dismissal had been based were simply too hasty.

Such work continues today, as cold fusion approaches its thirtieth birthday. A recent peer-reviewed Japanese paper lists seventeen scientific authors, from several major universities and the research division of Nissan Motors (Kitamura et al 2018). These authors report 'excess heat energy' which 'is impossible to attribute ... to any chemical reaction' (with good reproducibility between different laboratories). The field has also been attracting new investors recently (including, some claim, Bill Gates himself).

These seventeen Japanese scientists might be mistaken, of course. Scientists – not to mention investors! – often get things wrong. But their work is only the tip of a very substantial iceberg. If there was even a small probability that they and the rest of the iceberg were on to something, wouldn't the field deserve some serious attention, by the prudence principle with which we began?

When I wrote about these issues in *Aeon* three years ago, I argued that the problem is that cold fusion is stuck in a reputation trap. Its image is so bad that many scientists feel that they risk their own reputations if they are seen to be open-minded about it, let alone to support it. That's why the work of those Japanese scientists and others like them is ignored by mainstream science – and why it doesn't get the attention that simple prudence recommends.

The reputation trap is nicely illustrated by the tone of a *New Scientist* editorial from 2016. It accompanied a fairly even-handed article (Brooks, 2016) describing recent increases in interest in LENR, from investors as well as some scientists. The editorial concludes:

> There's still no compelling reason to think cold fusion will work. Let those with money to burn take the risk and, if proven right, the rewards for their chutzpah too. For the rest of us, cold fusion is better off left out in the cold. (Anonymous, 2016)

There's no mistaking the tone, but if we translate it to the safety case the logic has a chilling familiarity: 'There's no compelling reason to think that there will be icebergs at this latitude. Let those with money to burn take the slower route to the south, and the rewards if they turn out to be right.'

The fallacy here is obvious. It puts the burden of proof on the wrong side. What matters is not whether there is a compelling reason to think that there *are* icebergs, but whether there is compelling reason to be confident that there *are not*. That's what's distinctive about these safety cases, and it stems from the high cost of getting things wrong – hitting the icebergs, or missing the islands.

In the safety case, we know what happens when reputation and similar cultural and psychological factors get in the way of prudence. Icebergs are unlikely, and our reputation is at stake, so full speed ahead! NASA fell for precisely this trap in the case of the *Challenger* disaster, ignoring warnings about the O-rings (Berkes 2012). Something similar underlies the tone of the *New Scientist* editorial, in my view – a kind of misplaced rigidity, engendered in this case by the norms of scientific reputation.

Reputation plays an indispensable role in science, as an aid to quality control. But sometimes it gets things wrong. There are famous cases in the history of science in which new ideas were ignored or ridiculed, sometimes for decades, before going on to win Nobel prizes. (Classic examples include the work of Barbara McClintock on mobile elements in genetic material, and the discovery by Australian scientists Barry Marshall and Robin Warren that stomach ulcers are caused by a bacterium. )

Usually this doesn't matter very much – science got there in the end, in these famous cases. But it is easy to see how it might be a problem, where prudence requires that we take unlikely possibilities very seriously. If what's at stake is a serious risk, the normal rate of progress in

science – one funeral at a time, as Max Planck put it, commenting on science's conservatism – might simply be too slow.

So the normal sociology of scientific reputation may be pathological in special cases – cases in which the cost of wrongly dismissing a maverick idea view is especially high. In my *Aeon* piece I suggested that LENR is such a case. I proposed that to offset this pathology we need some carefully targeted incentives – an X-Prize for new energy technologies, say. To mainstream scientists this idea sounds absurd, even disreputable, at least in the case of cold fusion. But that's just the pathology talking, in my view – and the rational response to the pathology is to hack it and work around it, not to give way to it.

Not surprisingly, my article was controversial – some commentators wondered what it would do to my own reputation! Critics didn't disagree with the principle that we need to take low probability risks (or potential missed opportunities) seriously, when the cost of overlooking them would be high. But many denied that cold fusion falls into this category. They felt that it is so unlikely, so discreditable, that we can safely leave it in the reputation trap. Sometimes this response came with considerable vehemence, even from friends.

How likely would cold fusion have to be, to be worth serious attention? This is debatable, but a generous 5% should be uncontroversial. (Who would argue that we should ignore a 1 in 20 chance of some interesting new physics, let alone carbon-free energy?) My critics thought that the probability that cold fusion is real is much lower than that.

I felt that many of these critics were simply not paying attention. If one took the trouble to look, there was a lot of serious work, including recent work, suggesting real physical anomalies. If we ask not whether this evidence is entirely compelling, but simply whether it lifts the field above a very low attention threshold (say 5%), the answer seemed to me to be obvious. We shouldn't be ignoring this work. Instead, we should be trying to hack the pathology that makes it so easy to dismiss it.

In addition to scientists at respectable institutions who work on LENR, there are also some inventors and entrepreneurs who claim to be developing practical LENR-based devices. I mentioned two in my 2015 article. One was a controversial Italian engineer, Andrea Rossi. His claims in 2011 had attracted me to the topic in the first place, and in 2015 he seemed to be doing well. The other was a less colourful inventor, Robert Godes, whose Berkeley-based company Brillouin Energy also claimed to be on a path to a commercial LENR reactor.

My critics were confident that both Mr Rossi and Mr Godes must be frauds, or else deeply confused. What other possibilities are there, after all, if – as my critics were convinced – there's no genuine LENR? I thought that this dismissal was far too hasty. I wasn't certain by any means that Rossi or Godes did have what they claimed, but I thought that the probability was well above a reasonable attention threshold (given what success might mean).

With several critics, these differences of opinion led to bets, at long odds. I would win the bets if, after three years, at least one of Rossi or Godes has 'produced fairly convincing evidence (> 50% credence) that their new technology that generates substantial excess heat relative to electrical and chemical inputs.' If my opponents and I couldn't agree whether this is the case, the question would go to a panel of three judges for arbitration. Either way, the proceeds will support research on existential risk.

The three years is now up, so how am I faring? About Rossi, I am happy to concede that he hasn't made it to the finishing line, even at a modest 50% credence. I think there is still some reason to think that he may have *something,* based in part on claimed replications by far less colourful figures. But there is also evidence of dishonesty, especially in his dealings with his US backer, Industrial Heat.

Luckily for me, I backed the ants as well as the grasshopper. About Godes' Brillouin Energy (BEC) the news is much better. There are now three positive reports by an independent tester, Dr Francis Tanzella, at the Menlo Park lab of SRI International (Tanzella 2016, 2018a, 2018b). The first report already confirmed low levels of excess heat, and important progress in reproducibility:

> This transportable and reproducible reactor system is extremely important and extremely rare. These two characteristics, coupled with the ability to start and stop the reaction at will are, to my knowledge, unique in the LENR field to date.

The more recent reports describe steady progress in two directions. First, a modest improvement in excess heat as measured by the so-called Coefficient of Performance (COP) – the ratio of output power to input power. Second, a large increase in the *absolute* level of excess heat, from a few milliwatts in 2016 to several watts in early 2018.

The last of Dr Tanzella's three reports covers the period to July 2018. Since then, BEC themselves have claimed even better results – consistent output power around twice the level of input power, with excess heat of around 50 watts.

What are the options, if we are not to take these reports at face value? Essentially, one needs to dismiss as incompetent or fraudulent not only Mr Godes and his BEC team, but also Dr Tanzella and his SRI colleagues. However, as the 2016 report notes, SRI 'brought over 75 person-years of calorimeter design, operation, and analysis experience to this process' (Tanzella 2016, 1), much of it in the field of LENR. SRI, and Dr Tanzella himself, are among the most experienced experts in the field.

Accordingly, it seems to me greatly more likely than not that BEC do have what they claim – in the words of my bet, a device that 'generates substantial excess heat relative to electrical and chemical inputs'. Readers wishing to make up their own minds should study Dr Tanzella's reports, and listen to a recent podcast (Russell 2018) in which he speaks about his work. The same site also offers a recent interview with Robert Godes, in which he discusses BEC's latest results.

Some critics will say that Dr Tanzella must be wrong, because the claims are simply so unlikely. That would be an understandable view if BEC's claims were a complete outlier, unrelated to any previous work. But as I said, there's an iceberg's worth of work beneath it, much of it from eminently serious sources (people and institutions). Only someone who hadn't taken the trouble to look at this work could think of BEC as an outlier.

As a very small sample from this iceberg, see (McKubre 2009) and (Szpak et al 2008) for overviews of long programmes of work by two US laboratories, SRI International themselves and the Space and Naval Warfare (SPAWAR) lab, San Diego, over the 1990s and 2000s; see (Takahashi et al 2017) for a short summary of the recent Japanese work mentioned above; and see (Beiting 2017) and (Mizuno 2017) for two more recent technical papers. All these pieces report results not explicable by known chemical processes. The LENR-CANR.ORG site (LENR-CANR.ORG 2021) offers many hundreds of other papers.

Finally, for our Norwegian readers, there's a recent 45 page report by the Norwegian Defence Research Institute (Hasvold 2018). The author, an electrochemist, concludes that in his view 'LENR is a real phenomenon, the development of which ought to be closely watched.' He says that the alternative 'is to believe in a conspiracy of independent researchers at a number of different institutions', and adds that for the original Fleischmann and Pons reactions in particular, 'the documentation is highly convincing.'

The question I want you to ask yourself – *after* examining some of this material – is not whether you agree with me that BEC has made it over the finishing line specified in my bets. That's an interesting question, but not the important one. The crucial issue is whether LENR in general makes it over a much lower bar, the one that recommends it for serious attention, given how desperate we are for Bill Gates' energy miracle. If you don't agree with me even about the low bar, I'm wondering what you could possibly take yourself to know, that all these authors do not, that could justify such certainty?

If you do agree with me about the low bar, I encourage you to join me in trying to hack the reputation trap. It may be too much to expect mainstream science to scan the horizon very far to port and starboard. That's how science works, and rightly so, in normal circumstances. But if that's where the energy-rich islands might be, that's the direction someone needs to be looking. So we need some unconventional thinkers – especially young, brilliant, sharp-eyed thinkers – and we need to cheer not sneer at their efforts.

In my view, it's as much a mistake to let reputation blind us to prudence in this case as it was for the icebergs and O-rings. True, it isn't necessarily so catastrophic. But unlike the *Titanic* and the *Challenger*, the planet has all of us on board. So let's loosen our collars a little, remind ourselves of the virtues of epistemic humility, and do something to encourage our energy mavericks.

For the moment, as cold fusion turns thirty, it remains a black sheep of the scientific family. As the history of science shows us, however, it is often black sheep who bring home the black swans. We don't yet know whether cold fusion will follow the same course, but it is in everyone's interests to show it some warmth. For safety's sake, cold fusion needs to be cool.

## 7 Meeting McKubre

The piece above was published in early March 2019 (Price 2019). A few weeks later I met one of the central figures of the field, Dr Michael McKubre. McKubre, a New Zealand-born electrochemist, had worked with Martin Fleischmann in Southampton, some years before the 1989 announcement. He became a leading figure among the small group of scientists who regarded the dismissal of the Fleischmann and Pons results as too hasty. He was Director of the Energy Research Center at SRI International for many years.

By 2019 McKubre was semi-retired, spending most of the year in his home town of Napier, New Zealand. I was on leave in my home town of Sydney, Australia. I took advantage of our proximity to visit McKubre in Napier. We had dinner together on the thirtieth anniversary of the Pons and Fleischmann announcement in 1989. There were three thirtieth anniversary meetings that month: one at MIT, one at the Russian Academy of Natural Sciences, Moscow, and – most exclusive of all – McKubre and I, with our wives, in Napier, New Zealand.

McKubre gave me a fascinating insider's sense of the history of the field, and of the cold fusion landscape at that point. I was somewhat reassured to hear about the level of work that existed out of the public gaze. I recall particularly that McKubre said, without mentioning any names, that some of the well-known tech companies had quiet programmes.

## 8 Google Joins the Search

This claim was strikingly confirmed a couple of months later, when it became public that Google had been funding LENR work in several universities since 2014. To the surprise of many people, *Nature* published a Perspectives piece by some of the Google-funded researchers (Berlinguette et al 2019). McKubre himself later reported that he had had a hand in initiating that work in 2014, though he had not been involved in it since that time.

This is how the authors of the *Nature* piece present their work:

> The 1989 claim of 'cold fusion' was publicly heralded as the future of clean energy generation. However, subsequent failures to reproduce the effect heightened scepticism of this claim in the academic community, and effectively led to the disqualification of the subject from further study. Motivated by the possibility that such judgement might have been premature, we embarked on a multi-institution programme to re-evaluate cold fusion to a high standard of scientific rigour. Here we describe our efforts, which have yet to yield any evidence of such an effect. Nonetheless, a by-product of our investigations has been to provide new insights into highly hydrided metals and low-energy nuclear reactions, and we contend that there remains much interesting science to be done in this underexplored parameter space.
>
> So far, we have found no evidence of anomalous effects claimed by proponents of cold fusion that cannot otherwise be explained prosaically. However, our work illuminates the difficulties of producing the conditions under which cold fusion is hypothesized to exist. This result leaves open the possibility that the debunking of cold fusion in 1989 was perhaps premature because the relevant physical and material conditions had not (and indeed have not yet) been credibly realized and thoroughly investigated. Should the phenomenon happen to be real (itself an open question), there may be good technical reasons why proponents of cold fusion have struggled to detect anomalous effects reliably and reproducibly. Continued scepticism of cold fusion is justified, but we contend that additional investigation of the relevant conditions is required before the phenomenon can be ruled out entirely. (Berlinguette et al 2019)

Later in the piece, they conclude like this:

*Call to action*. Fusion stands out as a mechanism with enormous potential to affect how we generate energy. This opportunity has already mobilized a 25 billion dollar international investment to construct ITER. Simultaneous research into alternative forms of fusion, including cold fusion, might present solutions that require shorter timelines or less extensive infrastructure.

A reasonable criticism of our effort may be 'Why pursue cold fusion when it has not been proven to exist?'. One response is that evaluating cold fusion led our programme to study materials and phenomena that we otherwise might not have considered. We set out looking for cold fusion, and instead benefited contemporary research topics in unexpected ways.

A more direct response to this question, and the underlying motivation of our effort, is that our society is in urgent need of a clean energy breakthrough. Finding breakthroughs requires risk taking, and we contend that revisiting cold fusion is a risk worth taking. (Berlinguette et al 2019)

This message was very congenial indeed from my point of view, of course. But I would add two comments. First, *not revisiting cold fusion* is risk-taking, too, and potentially a much more serious one. That's the feature that this case shares with the more obvious cases of low probability high-impact risks – the high potential cost of a false negative.

Second, for the risks that (Berlinguette et al 2019) have in mind, the degree of risk depends on sociological factors. Do researchers put their own careers and reputations at risk? If so, we can do something about it, by pushing back against the reputation trap.

The motivation for this Google work had been much the kind of argument I made in my articles. (I'm not trying to claim any credit here. The Google programme pre-dates my first *Aeon* piece, and in any case, I take the point to be obvious to anyone not blinded by the reputation trap.) This was made clear in McKubre's own comment on the Google project, published a few months later.

> Two of the authors of the Perspective article, Ross Koningstein and David Fork, senior engineers at Google, previously wrote an article [Koningstein and Fork 2014] in which they analyze dispassionately earth's energy situation. In their vision, the known renewable energy sources and any conceivable [source], in their most optimistic projection, cannot supply the energy needs of our planet's growing and advancing population. One of their conclusions is that "new zero-carbon primary energy sources" must be developed. This article appeared in *IEEE Spectrum* in November 2014. Importantly, and before that, rather than congratulating themselves on their analysis and conclusions, the authors set out with Google's support to address that perceived need. The result is what we are discussing today, with the extension enumerated below. Google saw a problem, saw a potential solution, enlisted support and set out to do something about it. (McKubre 2019, 1)

What impact did the revelation of the Google work have on the reputation trap? Some in the LENR community felt that *Nature's* own initial reaction amounted to digging in. The same issue of *Nature* contained both an editorial and a commentary piece concerning the Google work. The latter was by the science writer Philip Ball, himself a former *Nature* editor, and was bylined like this:

> Why revisit long-discredited claims for a source of abundant energy, asks Philip Ball? Because we are still learning how to treat pathological science. (Ball 2019a, 601)

Clearly, that tells *Nature's* readers that cold fusion is pathological science, and is long-discredited. (Like 'refuted', 'discredited' is what philosophers call a success word – it takes sides on the facts.) As in the case of the *New Scientist* editorial I mentioned earlier, this byline has the grumpy tone of someone who hasn't yet come to terms with the fact that their earlier hostile judgement may have been too hasty.

Later in the piece, in Ball's own text, the message is qualified, though not by much:

> For some, cold fusion represented a classic example of pathological science. This term was coined in the 1950s to describe a striking claim that conflicts with previous experience, that is based on effects that are difficult to detect and that is defended against criticism by ad hoc excuses. In this view, cold fusion joins an insalubrious list that includes the N-rays of 1903, the polywater affair of the late 1960s and the memory of water episode of the late 1980s.
>
> *Nature* never published the manuscript by Fleischmann and Pons — the authors withdrew it to focus on follow-up work. But a paper reporting similar findings by a group at Brigham Young University in Provo, Utah, was published in April of that year (Jones 1989). The only report at the time from Fleischmann and Pons was a short paper, lacking in detail, in the *Journal of Electroanalytical Chemistry* (Fleischmann and Pons 1989).

> *Nature* did publish follow-up studies by other groups, including one that used the actual equipment of Fleischmann and Pons (Salamon *et al* 1990). None observed any hint of cold fusion, and no convincing evidence has since materialized. (Ball 2019a, 601)

Friends of LENR might respond that it is hardly surprising that none of the pieces published in *Nature* reported any hint of cold fusion, given the stance that *Nature* soon took on the subject. It was in the light of this well-known stance that the *Nature* Perspective piece in 2019 came as such a surprise. An effect of the stance had long been that positive results had to go elsewhere, to a few journals ready to take the reputational risk – a risk that *Nature* itself had done much to create.

Melinda Baldwin's excellent recent history of *Nature* gives this account of the episode.

> Although Pons and Fleischmann had indeed submitted an article to *Nature,* that journal never printed it: only Steven Jones's article, with its far more modest claims about neutron production and excess heat from the reaction, would be published in *Nature.* Instead of being the forum where a new era of energy was declared, *Nature* quickly became a major center of cold fusion skepticism. By 29 March 1990, a year to the week after the first mention of cold fusion in *Nature,* [the Editor John Maddox] felt secure enough to declare "Farewell (Not Fond) to Cold Fusion" in the magazine's leader.' (Baldwin 2015, 201)

As Baldwin goes on to say:

> During the cold fusion controversy, Maddox ... and the rest of the editorial staff cast the cold fusion episode as a battle between careful, peer-reviewed, properly conducted science and sloppy science revealed through press conferences in hopes of wealth through patents. Maddox wrote editorials criticizing Pons and Fleischmann's methods, associate editor David Lindley wrote news articles forecasting the death of cold fusion, and the journal's editorial staff gave significant space to cold fusion's most prominent scientific critics. Where *Nature* led, science reporters followed. News outlets such as *Time,* the *Economist,* and the *Wail Street Journal* all covered *Nature's* role in the cold fusion controversy and portrayed the journal's skepticism as proof that the scientific community was rejecting the Pons-Fleischmann claims. Ultimately, the cold fusion episode convinced many observers of the scientific journal's continued importance to the scientific community and illustrated *Nature's* influence among both scientists and laymen at the end of the twentieth century. (Baldwin 2015, 201–202)

Returning to Ball's commentary, the claim that 'no convincing evidence has since materialized' is clearly a judgement call, with which McKubre amongst others would simply disagree. Again, Ball chooses words that seem to leave no room for doubt on the matter.

A more important failing, in my view, is that nothing in this commentary acknowledges the issues of epistemic risk, to which the Google team are clearly sensitive – the potentially catastrophic cost of a false negative, in a case like cold fusion. *Nature* is one of the few institutions that could, if it chose, popularise this view, and hence do something about the reputation trap. Judged by this standard, I felt that Ball's commentary was a missed opportunity.

The editorial in the same issue does a little better. In this case the byline does not immediately close the door: 'Major project to reproduce controversial claims of bench-top nuclear fusion kindles debate about when high-risk research is worthwhile'; though again it misidentifies the important risk, which is that of *not* doing the research, given the high potential cost of a false negative. But the piece does go on to note that '[s]ociety's need for cheaper and cleaner sources of energy is more pressing than ever, and, if cold fusion were possible, it could be a disruptive technology with a world-changing pay-off.' (Anonymous 2019a)

The editorial concludes like this:

> The [Google] team found no evidence whatsoever of cold fusion.
>
> Is that the final nail in the cold-fusion coffin? Not quite. The group was unable to attain the material conditions speculated to be most conducive to cold fusion. Indeed, it seems extremely difficult to do so using current experimental set-ups — although the team hasn't excluded such a possibility. So the fusion trail, although cooling, is not yet cold, leaving a few straws for optimists to clutch on to.
>
> The question now is whether it is even worth continuing this research. Here, the message is more nuanced. The project has produced materials, tools and insights — such as calorimeters that operate reliably under extreme conditions, and techniques for producing and characterizing highly hydrided metals — that could benefit other areas of energy and fusion research. But whether the spin-off benefits alone justify continued efforts and investment in pursuit of a probable pipe dream is another matter. Opinions are split.
>
> So what do we take home from a multi-year failed experiment? First, that the programme has been conducted with rigour and attention to detail — we can have confidence in the results. Second, although the work provides no support for fringe groups that continue to insist that cold fusion exists, it does bring this research area back into the light of harsh scientific scrutiny. And, by doing so, the project might help responsible research in this general area to become less taboo, even if the chances of achieving cold fusion still look extremely remote. (Anonymous 2019a)

This doesn't miss the opportunity to wheel out some of the tropes of the reputation trap, and certainly ignores contrary evidence even from neutral expert assessors (more on this in §9.5). Some might feel that the remark about helping 'responsible research in this general area to become less taboo' is a little bit rich, coming from the journal that did so much to foster that view in the first place. But despite all that, the editorial does concede that the door is ajar.

A few months later, after meeting some of the Google team's researchers, Philip Ball published a second piece, this time in a journal with less of a horse in the race than *Nature* (Ball 2019b).[5]

> [C]old fusion has never gone away. A few researchers, working at the fringes of the scientific community, have continued to claim to see tantalizing signs that there really is something in it after all. However, the field has never shaken off its bad reputation. There was much surprise when in June, 30 years after the original event, *Nature* published an article by a team of researchers funded by Google describing renewed searches for "low-energy" fusion of hydrogen isotopes (deuterium, which has a lower energy threshold for fusion than hydrogen-1) using palladium electrodes.
>
> The paper reported no evidence of such a process in electrochemical experiments similar to those of Pons and Fleischmann, but it described a low level of fusion from a different experimental setup in which a plasma of deuterium ions surrounded a negatively charged palladium wire. The new findings will not persuade anyone that Pons and Fleischmann were right, but they could give cold fusion a new lease on life. Moreover, the study showed that there are interesting things still to learn about the materials science of the palladium–hydrogen system. (Ball 2019b, 883)

Most interestingly, Ball also spoke to some of the scientists involved in the Google-funded work. Here he quotes Curtis Berlinguette, head of the Berlinguette research group at UBC, Vancouver, and lead author on the *Nature* Perspectives piece.

> "Renewable energy and fusion technologies are not scaling at the pace we need them to," says Berlinguette. "If cold fusion were realizable, it could take the world into an era of energy surplus rather than scarcity. It therefore seemed irresponsible to not take another look at it. For me, cold fusion started in 2015," says Berlinguette. "Prior to that, I didn't know enough about it to have an opinion. I was driven simply by curiosity to learn more about the field."  (Ball 2019b, 883)

There is much to like here, from my point of view, especially Berlinguette's ability to look beyond the reputational factors to the true imperatives of the case ('It … seemed irresponsible not to take another look at it'). His epistemic attitude is also admirable. My experience in this field is that 'not knowing enough about it' doesn't prevent people from having very strong opinions – hence my plea for 'epistemic humility' at the end of my 2019 piece.

Berlinguette's webpage now says this: 'His program also likes to work on high risk, high impact clean energy projects like cold fusion' (UBC 2021). This is huge progress, in my view, in the sociological sense. In particular, it provides cover for younger scientists – the 'young, brilliant, sharp-eyed thinkers' I had in mind in §6* (Price 2019) – to work on these issues, without such a risk to their own careers.

---

[5] At around the same time, in October 2019, *Nature Materials* published an editorial that is clearly also informed by a good understanding of the Google team's work (Anonymous 2019b).

Has Google's own reputation provided cover, in this sense, for other research groups in the field? Not surprisingly, the answer is a resounding 'yes'. With this in mind, let's turn to other developments in the field since 2019.

## 9 The Global LENR Landscape – Updates since 2019

### 9.1 Japan

Let's begin with Japan. The Clean Planet initiative, first mentioned in my postscript from 2017 (§4*), continues to give the impression of steady progress, in two dimensions: both technological progress, and striking reputational progress, evident in Clean Planet's standing with significant organisations outside the LENR field.

Taking the latter dimension first, Clean Planet now has equity participation of two major Japanese companies. One is Mitsubishi Estate, a member of the Mitsubishi group. The second is Miura Co. Ltd, a major Japanese boiler manufacturer.

Miura invested in Clean Planet in May 2019. More recently, in September 2021, it signed an agreement with Clean Planet 'to jointly develop industrial boilers' (Miura 2021) based on Clean Planet's technology. This description of Clean Planet is from Miura's own press release about the collaboration. Note the reference to 'a major US IT company'.

> Clean Planet Inc. is a venture enterprise that has worked on the research and development of Quantum Hydrogen Energy, a safe, stable, and affordable source of clean energy, in order to create groundbreaking innovations in the energy industry, a vital element of social infrastructure. Quantum Hydrogen Energy is currently attracting attention globally, and large companies and investors representing every industry are beginning full-scale participation in this field, as can be seen from the entry of a major US IT company. Against this background, Clean Planet Inc. has embarked on a range of cutting-edge research and development efforts in collaboration with Tohoku University, and in April 2021 began work on development for practical use for release of one kilowatt of thermal energy using Quantum Hydrogen Energy. (Miura 2021)

This news has received some coverage in the mainstream Japanese business press, though not as far as I know in the West. This is from the NikkeiBP website:

> Focusing on [Clean Planet's] research results, Mitsubishi Estate invested in Clean Planet in January 2019 and Miura Co., Ltd. invested in Clean Planet in May of the same year. Since then, research has progressed steadily toward practical use, so [Clean Planet] decided to start full-scale joint development with Miura Co., Ltd. regarding its application to industrial boilers. A prototype will be produced in 2022 and will be commercialized in 2023. (NikkeiBP 2021)

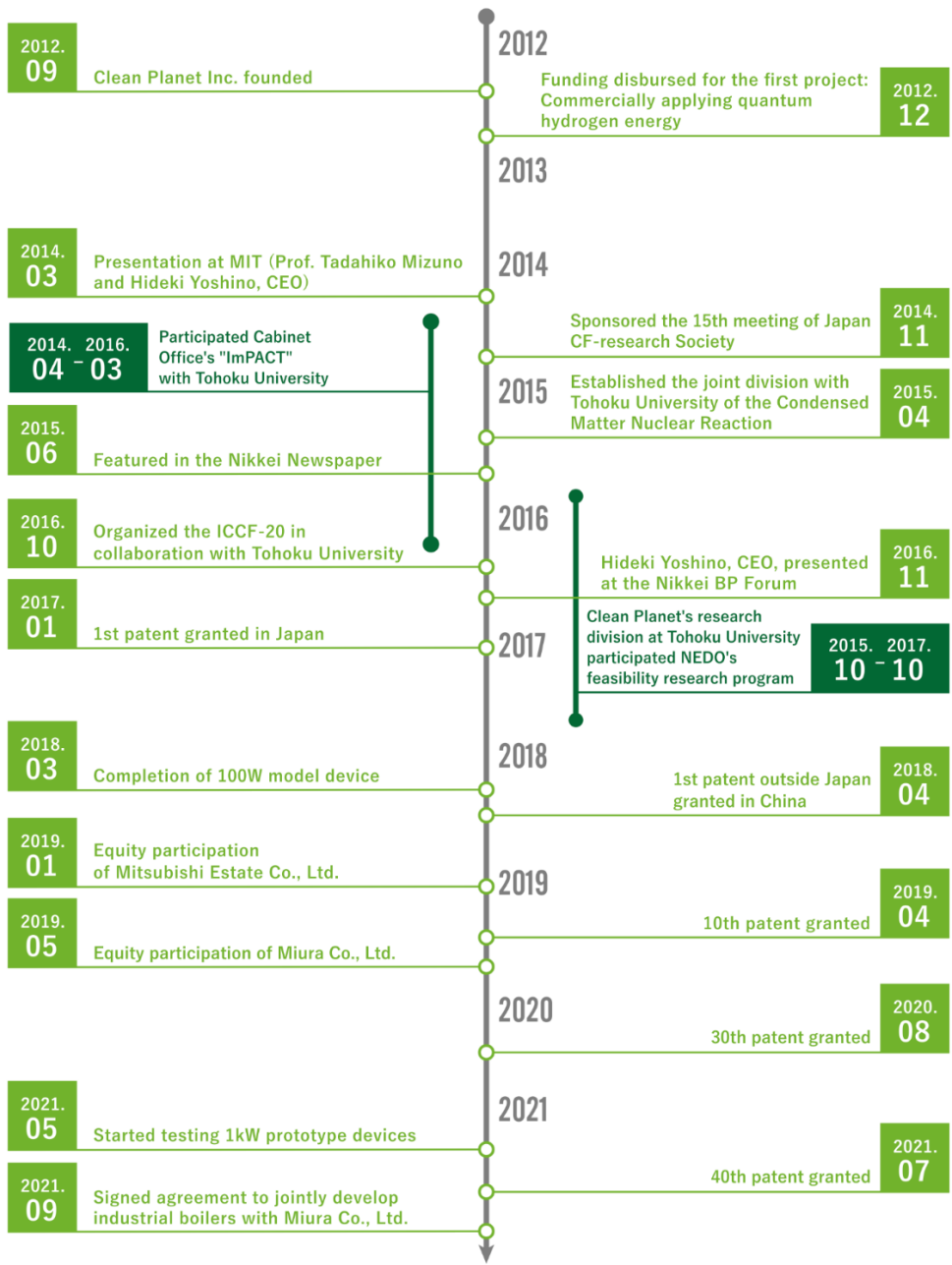

Figure 1: Clean Planet timeline (from Clean Planet 2021)

An excellent impression of Clean Planet's apparent progress over the past decade is conveyed by Figure 1, reproduced with permission from the Clean Planet website (Clean Planet 2021). Note, in particular, the reported timings of completion of a 100W 'model device' (2018) and 'starting testing' of 1kW prototypes (May 2021).

I would like to encourage readers to consider the possible explanations for the existence of this diagram, and the claims it embodies. So far as I can see, there are essentially three possibilities:

(i) We take the claims at face value, and accept that Clean Planet does have devices producing large amounts of heat, not explicable in chemical or other nonnuclear terms.

This means accepting that so-called cold fusion is real (even if perhaps badly named – this semantic point would need to await a good theoretical explanation of the excess heat in question).

(ii) We accept that the claims are sincerely made, but take them to result from gross error of some kind – either measurement error, or failure to spot an alternative explanation for the results, not requiring LENR.

(iii) We deny that the claims are sincerely made, but take them to result from deception or misrepresentation of some kind (either by Clean Planet personnel themselves, or by someone else – perhaps the entire website is a hoax by high school hackers, for example).

Unless there is some further possibility that I have missed, it is a rational requirement that you divide your degrees of credence between the three possibilities (i)–(iii) in a way that adds up to 1, or 100%. (If you don't do that, a clever Dutch bookie can make a combination of bets with you that you are guaranteed to lose, no matter what – see Vineberg 2016.) I encourage readers who continue to give negligible credence to option (i), the reality of cold fusion, to consider what credence they wish to give to the other two possibilities. Please don't dodge the question by saying that you don't have access to the evidence. Take the diagram itself as your evidence. You do have access to the diagram – that's why I have presented the options in this way. Ask yourself what credence you give to the three possible explanations for *its* existence.

What do I think? In my view option (iii) is vanishingly unlikely, in this case. Gross error is possible, but in my view now rather unlikely – even more so than in 2019, given the reported progress concerning the size of the effect. So I now give less than 20% credence to option (ii). Accordingly, I give at least 80% credence to the option (i), that of taking the claims at face value.

As I said with respect to Brillouin in §6* (Price 2019), resistance to this conclusion would be understandable if Clean Planet's claims were complete outliers, unrelated to any previous scientific work. But they're not. Like Brillouin, Clean Planet is simply one of the most prominent tips of a substantial iceberg of scientific work, over thirty years.

Readers may also feel sceptical on the grounds that if there were something to Clean Planet's claims, we would have heard a lot more about them. Where was the news of Clean Planet's collaboration with Miura in the month of COP26 in Glasgow, for example? But you know the answer to this. The reputation trap continues to ensure that LENR remains unmentionable, by most journalists. Would you put your head above the parapet, and be the first major science writer to be seen to be taking it seriously?

**9.2 Europe**

Turning now to Europe, I know of no recent scientific or technological progress to compare with that in Japan. But the sociological dimension is a different matter. The EU's flagship Horizon 2020 funding programme recently awarded two large grants to multi-institution teams working on LENR. This is a very significant shift, in my view.

The first of these teams, the HERMES project, has been funded at €3,999,870 over the five years 2020–2024. Coordinated from Finland, HERMES is described like this on the funder's website. The leverage provided by the Google work is again explicit.

> In 1989, electrochemists Martin Fleischmann and Stanley Pons made headlines with their claim to have produced excess heat using a simple apparatus working at room temperature. Their experiment involved loading deuterium in a palladium metal. As many experimenters failed to replicate their work, cold fusion remains a controversial topic in the scientific community. Nevertheless, a vociferous minority still believes in this elusive phenomenon. Since 2015, Google has been funding experiments into cold fusion. Although no evidence has been found for this phenomenon, it is clear that much pioneering research remains to be conducted in this poorly explored field. The EU-funded HERMES project will employ advanced techniques and tools developed over the last few decades to investigate anomalous effects of deuterium-loaded palladium at room and intermediate temperatures. (CORDIS 2021a)

The second EU grant, from the same Horizon 2020 funding programme, is for the CleanHME project. This was funded at €5,678,597.50 over 2020–2024. It is coordinated from Poland, and described like this:

> With climate change being a major global concern in recent times, new efficient clean energy sources are in high demand, and there has been a rise in the use of many of them, such as solar or wind generators. One very promising energy source is hydrogen–metal energy (HME), which could be used for small mobile systems as well as in stand-alone heat and electricity generators. Unfortunately, little research has been conducted concerning HME. The EU-funded CleanHME project aims to change this. It will produce an elaborate comprehensive theory of HME phenomena that would assist in the optimisation of the process and construct a compact reactor to test HME technology. (CORDIS 2021b)

Again, these grants indicate a very welcome change in the sociological climate, in my view.

**9.3 USA – new results from a longstanding NASA programme**

Moving now to the US, our first stop is at NASA. There has been some interest in LENR within NASA since at least the early 2010s. For example, (Wells et al 2014) reports on a NASA-funded project investigating the potential of LENR as a power source for terrestrial flight, while (Bushnell et al 2021) is a NASA report on its potential for space flight. But a much more hands-on approach to the subject came to light in 2020, as a NASA website reports:

> *NASA Detects Lattice Confinement Fusion.* A team of NASA researchers seeking a new energy source for deep-space exploration missions, recently revealed a method for triggering nuclear fusion in the space between the atoms of a metal solid. Their research was published in two peer-reviewed papers in the top journal in the field, *Physical Review C* [Pines et al 2020; Steinetz et al 2020] ...

Nuclear fusion is a process that produces energy when two nuclei join to form a heavier nucleus. "Scientists are interested in fusion, because it could generate enormous amounts of energy without creating long-lasting radioactive byproducts," said Theresa Benyo, Ph.D., of NASA's Glenn Research Center. "However, conventional fusion reactions are difficult to achieve and sustain because they rely on temperatures so extreme to overcome the strong electrostatic repulsion between positively charged nuclei that the process has been impractical."

Called Lattice Confinement Fusion, the method NASA revealed accomplishes fusion reactions with the fuel (deuterium, a widely available non-radioactive hydrogen isotope composed of a proton, neutron, and electron, and denoted "D") confined in the space between the atoms of a metal solid. In previous fusion research such as inertial confinement fusion, fuel (such as deuterium/tritium) is compressed to extremely high levels but for only a short, nano-second period of time, when fusion can occur. In magnetic confinement fusion, the fuel is heated in a plasma to temperatures much higher than those at the center of the Sun. In the new method, conditions sufficient for fusion are created in the confines of the metal lattice that is held at ambient temperature. While the metal lattice, loaded with deuterium fuel, may initially appear to be at room temperature, the new method creates an energetic environment inside the lattice where individual atoms achieve equivalent fusion-level kinetic energies. (NASA Glenn Research Center 2021)

We learn from this that the NASA team are as skilled at splitting hairs as they are at physics. (Koziol 2020) picks up this aspect of the story, at *IEEE Spectrum:*

"What we did was *not* cold fusion," says Lawrence Forsley, a senior lead experimental physicist for the project. Cold fusion, the idea that fusion can occur at relatively low energies in room-temperature materials, is viewed with skepticism by the vast majority of physicists. Forsley stresses this is hot fusion, but "We've come up with a new way of driving it."

"Lattice confinement fusion initially has lower temperatures and pressures" than something like a tokamak, says [Theresa Benyo, an analytical physicist and nuclear diagnostics lead on the project]. But "where the actual deuteron-deuteron fusion takes place is in these very hot, energetic locations." Benyo says that when she would handle samples after an experiment, they were very warm. That warmth is partially from the fusion, but the energetic photons initiating the process also contribute heat. (Koziol 2020)

It is not clear to me that the notion of 'locally hot' makes sense at this scale. Temperature is a statistical property, defined in terms of the mean kinetic energy of the atoms or molecules of a substance. For individual molecules, atoms or subatomic particles, it is more appropriate to speak of their energy than their temperature.

In any case, these remarks are clearly an attempt by the NASA group to protect themselves from the heat of the reputation trap. Forsley himself has long been a significant member of the cold fusion community, appearing with other major figures in the field as co-author of (Marwan et al 2010), for example. There is an excellent interview with him, discussing this recent NASA work, at (Hughes 2020).

One imagines the ghost of Martin Fleischmann rolling his eyes, and pointing out that what the NASA group have in mind is perfectly compatible with his own use of the term cold fusion, which never referred to anything more than the temperature of the test tube – and then being ticked off by sceptics, keen to cover their own retreat, for 'not making that clear in the first place'![6]

### 9.4 New naval manoeuvres

Turning from deep space to deep sea, the US Navy has long supported research on cold fusion. In §6* (Price 2019) I mentioned work by the Space and Naval Warfare (SPAWAR) lab in San Diego. The combination of the NASA news and the Google news apparently inspired a different Navy group to pull together a collaboration of US Government labs, as another *IEEE Spectrum* piece describes.

> After more than three decades of simmering debate in specialized physics groups and fringe research circles, the controversy over cold fusion (sometimes called low-energy nuclear reactions or LENRs) refuses to go away. On one hand, ardent supporters have lacked the consistent, reproducible results and the theoretical underpinning needed to court mainstream acceptance. On the other, vehement detractors cannot fully ignore the anomalous results that have continued to crop up, like the evidence for so-called "lattice-confinement fusion" adduced last year by a group at NASA's Glenn Research Center.
>
> Scientists at the Naval Surface Warfare Center, Indian Head Division have pulled together a group of Navy, Army, and National Institute of Standards and Technology (NIST) labs to try and settle the debate. Together, the labs will conduct experiments in an effort to establish if there's really something to the cold fusion idea, if it's just odd chemical interactions, or if some other phenomenon entirely is taking place in these controversial experiments.
>
> In 1989, electrochemist Stanley Pons and chemist Martin Fleischmann published the results of experiments in which they claimed to observe anomalous heat, as well as fusion by-products like neutrons, in a simple, room-temperature tabletop set-up involving palladium and heavy water. The claim was, to put it mildly, huge. Fusion is typically a high temperature, high pressure phenomenon. It requires a star, or, if you're hellbent on making it happen on Earth, massive magnets and a lot of power. Yet the promise of cheap, safe, and abundant energy was soon dashed when the vast majority of scientists failed to replicate their results.

---

[6] I believe that Fleischmann and Pons did not use the term 'cold fusion' themselves, initially, though it was soon used to describe their claims.

But still, lingering interesting results continued to emerge. Aside from the recent promising findings from NASA, Google published a paper in *Nature* in 2019 revealing that the company had spent US $10 million to research cold fusion since 2015. The company teamed up with researchers at institutions including MIT, the University of British Columbia, and Lawrence Berkeley National Laboratory. The research group found no evidence of classic Pons-Fleischmann-style cold fusion, but it did find evidence of the larger umbrella category of LENRs—suggesting (as the NASA group also reported) that nuclear fusion may be possible in locally-hot sites in otherwise room temperature metals.

"We got our impetus from the Google paper appearing in *Nature*," says Carl Gotzmer, Indian Head's Chief Scientist. Gotzmer's duties include keeping the Navy abreast of the latest scientific developments. Gotzmer says his cold fusion/LENR interest developed after attending the International Conference on Cold Fusion in 2003. After a four-hour conversation with Fleischmann himself, and seeing presentations from across the world giving evidence of nuclear transmutations, he says he began to follow this field in earnest.

"Quite frankly, [to] other folks who have tried this over the years, it was considered a career ender," says Gotzmer. But the Indian Head team decided that, as a government lab, they had a little more freedom to pursue a controversial topic, so long as it also offered up the prospect of rewarding scientific results.

"I'm not as worried about looking into something that is considered controversial as long as there's good science there," says Oliver Barham, a project manager at Indian Head involved in the effort. "The whole point of our effort is we want to be doing good science. We're not out to prove or disprove anything, we're out to assemble a team of scientists who want to take it seriously." (Koziol 2021)

**9.5 The ARPA-E workshop (October 2021)**

One of the most significant actors in the story of cold fusion is the US Department of Energy (DOE). DOE produced reports on the field in 1989 and 2004 – more on those below. The Advanced Research Projects Agency-Energy (ARPA-E) is the DOE's version of the well-known Defense Advanced Research Projects Agency (DARPA) – the agency 'credited with such innovations as GPS, the stealth fighter, and computer networking', as the ARPA-E website puts it. ARPA-E itself says that it 'advances high-potential, high-impact energy technologies that are too early for private-sector investment'. (ARPA-E 2021a)

In October 2021, ARPA-E hosted a publicly announced workshop on LENR. This is how it is described on their own website:

> The objective of this workshop was to explore compelling R&D opportunities in Low-Energy Nuclear Reactions (LENR), in support of developing metrics for a potential ARPA-E R&D program in LENR. Despite a large body of empirical evidence for LENR that has been reported internationally over the past 30+ years in both published and unpublished materials, as well as multiple books, there still does not exist a widely accepted, on-demand, repeatable LENR experiment nor a sound theoretical basis. This has led to a stalemate where adequate funding is not accessible to establish irrefutable evidence and understanding of LENR, and lack of the latter precludes the field from accessing adequate funding. Building on and leveraging the most promising recent developments in LENR research, ARPA-E envisions a potential two-phase approach toward breaking this stalemate: (1) Support targeted R&D toward establishing at least one on-demand, repeatable LENR experiment with diagnostic evidence that is convincing to the wider scientific community (focus of this workshop); (2) If phase 1 above is successful (metrics to be determined), support a broader range of R&D activities (to be defined later) toward better understanding of LENR and its potential for scale-up toward disruptive energy applications, thus setting up LENR for broader and more systematic support by both the public and private sectors. (ARPA-E 2021b)

The website adds that for the purposes of the workshop, 'LENR is defined as a not-yet-understood process (or class of processes) characterized by system energy outputs characteristic of nuclear physics (typically >> 1 keV/amu/reaction) and energy inputs characteristic of chemistry (~eV/atom)'.

The website now includes links to the workshop presentations from a number of speakers, including several who will be familiar to readers who have made it this far: Michael McKubre, the NASA team, Clean Planet, and Brillouin, for example. While it is too soon to know what the impact of this workshop will be, it does seem to represent a further welcome shift in the sociological climate. Concerning the physics, I particularly recommend the presentation by Florian Metzler (MIT), which does a very interesting job of pulling together various strands in the LENR literature (Metzler 2021).

There is also a presentation from ARPA-E Fellow, Dr Katherine Greco, about the reviews of the field conducted by DOE in 1989 and 2004. Although the LENR community was disappointed by the 2004 review, it was certainly not wholly negative. As Greco's presentation puts it, the DOE panel was 'nearly unanimous' that

> funding agencies should entertain individual, well designed-proposals [into]
> – Whether or not there is anomalous energy production in Pd/D systems
> – Whether or not D-D fusion reactions occur at energies ~eV (Greco 2021)

Some LENR researchers complain that this recommendation was not actually followed – that funding proved as elusive as before (Maguire 2014). For my purposes, however, what's relevant is that DOE's panel of 18 rather sceptical physicists did not simply dismiss the field. Their individual views varied. McKubre (Maguire 2014) reports that the nine who participated in an in-person one-day meeting were almost uniformly positive. The report itself states that on the question of 'experimental evidence for the occurrences of nuclear reactions in condensed matter at low energies (less than a few electron volts)', one third of the reviewers were either completely convinced or 'somewhat convinced' (DOE 2004). And collectively, in any case, they left the door ajar. In contrast to *Nature* much more recently, the DOE reviewers did *not* treat LENR as 'long-discredited … pathological science' (Ball 2019a).

Meanwhile, DOE itself has just announced plans to establish a new Office of Clean Energy Demonstrations.

> President Biden's Bipartisan Infrastructure Law provides more than $20 billion to establish the Office of Clean Energy Demonstrations and support clean energy technology demonstration projects in areas including clean hydrogen, carbon capture, grid-scale energy storage, small modular reactors, and more. Demonstration projects prove the effectiveness of innovative technologies in real-world conditions at scale in order to pave the way towards widespread adoption and deployment. The founding of this office represents a new chapter that builds on DOE's long-standing position as the premier international driver for clean energy research and development, expanding DOE's scope to fill a critical innovation gap on the path to net-zero emissions by 2050. (DOE 2021)

In the light of DOE's reluctance to fund LENR research in 1989 and (as it has turned out) since 2004, it will be interesting to see whether it is now actually moving in a different direction. Certainly the emphasis here on 'demonstrations' chimes very nicely with what the ARPA-E workshop had in mind.

**9.6 The US Defence Intelligence Agency (2009)**

To balance this impression of *new* interest, let me end this list of developments involving US Government agencies with newly-available evidence of much earlier interest. In response to a freedom of information request, the US Defence Intelligence Agency has recently released a 2009 report titled 'Technology Forecast: Worldwide Research on Low-Energy Nuclear Reactions Increasing and Gaining Acceptance' (DIA 2009).

This report includes a detailed list of work undertaken in the period 1989–2009, in many parts of the world. It then concludes like this:

> Although no one theory currently exists to explain all the observed LENR phenomena, some scientists now believe these nuclear reactions may be small-scale deuterium fusion occurring in a palladium metal lattice. Some others still believe the heat evolution can be explained by non-nuclear means. Another possibility is that LENR may involve an intricate combination of fusion and fission triggered by unique chemical and physical configurations on a nanoscale level. **This body of research has produced evidence**

> **that nuclear reactions may be occurring under conditions not previously believed possible.** (DIA 2009, emphasis in the original)

The report goes on to discuss potential applications of LENR, 'if nuclear reactions in LENR experiments are real and controllable'. When I read it, I recalled the scientists I mentioned in my original *Aeon* piece, who were technology forecasters for a similar UK organisation – and who told me that they had trouble getting their agency to take LENR seriously, even as a low probability possibility. It turns out the work they needed had been done by some of their US counterparts, several years earlier.

**10 Summary – the State of the Field**

In my original *Aeon* piece (Price 2015) I expressed this view:

> I suspect it's too late to dismantle the [reputation] trap for LENR – the horse is already in the process of bolting, I think. If so, then the field is going to be mainstream soon, in any case. But we could try to learn from our mistakes. There may be other potential cases with a similar payoff structure (a high cost for false negatives, with a low cost for false positives).

This now seems to me to have been mistaken in one respect: I chose the wrong creature, or at least the wrong gait. In the light of the rate of progress I hoped for then, LENR has been more of a stroll than a gallop. Still, we know what slow and steady does, and this is still what I expect for LENR. That is, I expect that the science of LENR (defined, let us say, in the terms specified by the recent ARPA-E workshop) will indeed become mainstream. It is too soon to tell whether it will turn out to be useful as a source of energy, but I think the present signs are quite encouraging about that, too.

As I have noted at several points, it is not unusual in the history of science for formerly controversial ideas to become mainstream. So if LENR goes the way I expect, it will not be exceptional on those grounds alone. If there is a case for regarding it as exceptional, in the long run, I think it will rest on two main factors.

The first factor will be the depth of the reputational trap from which it will have managed to dig itself out. Here I'm thinking of several things: the severity of the condemnation and ridicule to which the field has been subject; its highly public nature; and the involvement of major scientific institutions such as *Nature* in administering it. I stress that I'm not making a critical judgement at this point about people or institutions who played a part in giving the field this reputation; I'm aware, of course, that there's a case to be made that Fleischmann and Pons set themselves up for it, in choosing their own public and unconventional course.

Where I do make a critical judgement concerns the second factor that will make LENR exceptional, if it does return to the mainstream. This is the point I have stressed from the beginning, about the high cost of a false negative. I mentioned Heather Douglas, who spoke at our maverick workshop at Trinity College in 2017. This is from a piece by Douglas called 'Rejecting the Ideal of Value-Free Science' (Douglas 2007).

> In general, if there is widely recognized uncertainty and thus a significant chance of error, we hold people responsible for considering the consequences of error as part of their decision-making process. Although the error rates may be the same in two contexts, if the consequences of error are serious in one case and trivial in the other, we expect decisions to be different. Thus the emergency room avoids as much as possible any false negatives with respect to potential heart attack victims, accepting a very high rate of false positives in the process. … In contrast, the justice system attempts to avoid false positives, accepting some rate of false negatives in the process. Even in less institutional settings, we expect people to consider the consequences of error, hence the existence of reckless endangerment or reckless driving charges.

Douglas goes on to discuss the possibility that '[w]e might decide to isolate scientists from having to think about the consequences of their errors', but rejects it. She argues that 'we want to hold scientists to the same standards as everyone else', and therefore 'that scientists should think about the potential consequences of error.' (Douglas 2007)

In my view, the blameworthy feature of the cold fusion case, if it becomes mainstream – indeed, I think it is blameworthy, *whether or not* it becomes mainstream – is the apparent failure of many of its critics to take this 'Douglas Doctrine' into account. Of course, it is too soon to judge whether this has made any practical difference. We don't yet know whether LENR will turn out to be a useful energy source. Even if so, it will be difficult to estimate how long a delay the treatment of the field might have caused.

But these unknowns are in one sense irrelevant. We don't excuse lax safety practices just because a disaster fails to happen as a result. Think of Douglas's examples – reckless endangerment and reckless driving. A person may be guilty of these things, even if by good fortune they fail to harm anyone. I think there's a prima facie case for charging some of the institutions of science with reckless endangerment, or something on that spectrum, in the case of LENR.

However, it would take a much better historian than me to write a detailed brief for such a charge, or indeed to try to make the case for the defence. Did actors such as John Maddox at *Nature* ever give consideration to the costs of a false negative, for example? Rather than trying to answer that historical question, let me turn to a more useful one. How can we do better in future cases with a similar risk profile? I have several suggestions.

**11 Recommendations**

As I said at the beginning, much of my interest in the case of cold fusion stems from its similarities with other cases in which hasty dismissal of an unconventional scientific claim might be dangerous. In particular, I have in mind the cases in which the claim in question concerns a potential catastrophic risk. My recommendations here are offered with all of these cases in mind.

## 11.1 Foregrounding the Douglas Doctrine

The most obvious recommendation is that the factors to which Heather Douglas calls attention need to be better known. They need to be internalised both by scientists themselves, and by important scientific institutions, such as major journals and learned societies. As Douglas's examples show, the points themselves are not difficult to see, and are already built into standard practice in many contexts, such as emergency rooms and law courts. But they need to be much more familiar, and to be encapsulated in simple maxims that come to seem a matter of common sense.

By way of comparison, think of the principle that correlation is not causation. That rolls so easily off the tongue, these days, that no one has any excuse for ignoring it. Imagine the incredulous response to someone who does so: 'You didn't *realise* that correlation need not imply causation? Where have you been?'

If we had one or two similarly familiar phrases capturing the principle that tolerance for error needs to depend on what's at stake, it would be harder for anyone to ignore them. I've called it the Douglas Doctrine in the hope that in this case, too, some catchy alliteration will help. But a catchy descriptive version would be even better. In §6* (Price 2019) I put it this way: 'The more disastrous a potential failure, the more improbable it needs to be before we can safely ignore it.' Perhaps 'High stakes/low bar' might serve as an easily memorable version of this principle?

## 11.2 Understanding reputation traps

My second recommendation is that scientists and scientific institutions need a better understanding of the way in which reputation operates in science, especially negative reputation. Hopefully, such an understanding would provide a degree of self-awareness, a willingness to consider whether one's own reactions are to be trusted. The pathology of the reputation trap, as I've called it, is that reputational factors get in the way of listening to the 'low probability voices', to the mavericks who need to be heard when a lot is at stake. If this were well understood, it would be much easier to take steps to avoid it.

Obviously, this recommendation combines with the first. The high stakes/low bar principle might tell us that a fringe view needs to be taken seriously; while an understanding of the grip of the reputation trap might make that easier.

## 11.3 Improving the climate of scientific debate

Understanding the pathology of reputation traps is one thing, but my third recommendation is we do something to discourage them in the first place. I think we need to pay attention to the language and 'climate' of scientific debate. In particular, we need to be conscious of the role of what we can call *epistemic slurs*. I use this term in the sense of the philosopher Josh Habgood-Coote, who argues that we should 'stop talking about "fake news"' (Habgood-Coote 2018, 2019):

> 'Fake news' has a rich expressive meaning, and often functions as an *epistemic slur*. Applying 'fake news' to a news story seems not to describe the story, but to express disdain toward the story, the institution that produced it, and (in some cases) toward people that believe the story.

In my case, I have in mind terms such as 'crank', 'crackpot', 'chicanery', 'pseudoscience', 'conspiracy theory', 'debunk', and the like. These terms do have some descriptive content, to varying degrees, but all of them also function as insults. They 'express disdain' towards a person or a view, as Habgood-Coote puts it. In other words, the target is accused not merely of an *epistemic* deficit – e.g., of making claims that are insufficiently supported by evidence – but of what we might call a connative deficit, as well. 'He's not just mistaken, he's a crank!'

When used effectively, these terms thus function to put their targets in categories defined by our disdain towards their members. It is worth asking whether this is necessary, or helpful, in scientific debates. After all, one obvious consequence of the availability of such labels is that they are easily used by others, including many who are not competent to participate in the scientific debate themselves.

A possible remedy would be to attempt to flag and deprecate the use of such epistemic slurs – to make them seem unacceptable, or at least less acceptable, in scientific argument. We already do this with slurs of other kinds, in other contexts – think of prohibitions on so-called hate-speech. Many workplaces have guidelines in place to discourage verbal bullying (as well as other kinds of bullying).

In my own discipline, there has been a recent movement to improve the culture of discussion in Q&A after research talks and similar events; see (NYU Philosophy 2021), for example, for the kind of guidelines that are now common. This has been a response to the complaint that many people found some aspects of the old culture hostile and intimidating. Something similar could be done in science, I think. We would end up presenting a less hostile face to views with which we disagreed, because some of the ways of signalling hostility had been discouraged.

It might be objected that such a move would make it harder to police the boundaries of science, to separate good science from bad. But if so, in my view, it is not clear that that would be a bad thing. Think of slurs as a form of verbal violence. Real police officers no doubt find it easier to police boundaries (e.g., to separate protestors from members of the public) if they are allowed to resort to violence, but that doesn't mean it's a good thing. Separating good science from bad certainly matters, but that doesn't mean that 'anything goes' in the attempt to do so.

Like physical violence, slurs also make it easier for people and institutions to erect barriers in places that suit their own (non-epistemic) interests. If cold fusion had turned out to be 'the real deal' in 1989, many other research programmes and economic activities would have been massively affected. Did such considerations ever get in the way of a fair hearing for cold fusion? That's another historical question I'll leave to one side, but it would hardly be surprising if factors like these had played a role. Think of the (now) well-known opposition to climate science over recent decades, e.g., by fossil fuel companies.

My point here is a different one. Epistemic slurs provide a very powerful weapon for actors who are not motivated simply by getting the science right, enabling them to attack the *reputation* rather than the *arguments* of their opponents. This is a version of the famous *ad hominem* fallacy (Wikipedia 2021d), of course. Like that fallacy in other contexts, it is often a sadly effective rhetorical device. In my view, it would be very much in the interests of good science to lessen this risk, by shining a disapproving light on the language that facilitates it.

Clearly, this point is very general. Many commentators have noted the degeneration in the climate of public scientific debate, especially in the Covid-19 epidemic. If I have a novel claim, it is only a reminder that these issues take on a particular urgency in cases in which the costs of wrongful dismissal are exceptionally high. In other words, one corollary of the Douglas Doctrine is that we need to be especially careful of epistemic slurs in cases with this value profile.

**11.4 Epistemic humility**

At the end of §6* (Price 2019) I recommended that we should 'loosen our collars a little, remind ourselves of the virtues of epistemic humility, and do something to encourage our energy mavericks.' What did I have in mind by the phrase 'epistemic humility'? A couple of things, one general and one particular.

The general point is that if we want to pay attention to low-probability options, then we need to leave the door ajar. We can't afford to take mainstream confidence that those views are mistaken as a reason for excluding them from the scientific conversation. Accordingly, an appreciation that one's own strongly-held beliefs might be mistaken is an important virtue in this kind of context. This is the main thing I had in mind by epistemic humility; see (Angner 2020) for a similar plea for epistemic humility in the Covid-19 pandemic.

The particular point I had in mind was a view about some of my friends and discussants, much more sceptical than me about cold fusion. I felt that they were insufficiently attentive to the evidence provided by better-placed observers, and I regarded this as a failure of epistemic humility. In (Price 2016) I expressed the thought like this:

> My recommendation to these friends is to keep at least one eye open. For herd animals – like ostriches and scientists – a good way to know when to move is to keep an eye on peers who are closer to the action. If they start to shift, then you should consider it too – unless you have good reason to think that you know something that they don't.

I returned to this kind of thought at the end of §6* (Price 2019). I list a number of publications by (apparently) well-qualified authors, all of whom clearly believe that LENR should be taken seriously, and that they themselves have the evidence to support such a view. I ask the reader whether they feel that this material lifts the subject above the very low probability bar that would recommend LENR for serious attention, in the light of the high impact/low bar principle. I say:

> If you don't agree with me even about the low bar, I'm wondering what you could possibly take yourself to know, that all these authors do not, that could justify such certainty?

In §6* I suggest setting the bar at 5%, which I felt was generous to my opponents. (A probability of 5% would justify an investigation for interesting new physics, even if it didn't have potential implications for the energy crisis.) Yet some of my critics, not physicists themselves, felt that LENR didn't get anywhere near this bar. As I say, I felt that their confidence that their own view trumped that of better-qualified observers closer to the action showed a regrettable lack of epistemic humility.

I now realise that to the evidence offered in §6* I could have added the 2004 DOE report, mentioned above. While not seen by insiders as friendly to the field, this report certainly wasn't as dismissive as the critics I have in mind. Indeed, more than 5% of the DOE assessors (i.e., one out of 18) found the evidence for LENR *completely* convincing, and a third of them found it at least partially so. These were experts called in by DOE to examine evidence and give an opinion. Some of my critics seemed to feel that they could do better from their armchairs.

Some of these critics were linked to the Silicon Valley rationalist community, a group with an admirable commitment to epistemic self-improvement (Bay Area Rationalists 2021). Because they lived nearby, I suggested that I could introduce them to my LENR contacts in Silicon Valley – e.g., to Francis Tanzella at SRI International, or (subject to signing an NDA) to the Brillouin team. But this seemed to elicit no interest. I was reminded of a remark from Brillouin's Robert Godes, quoted in §2* (Price 2015): 'It is sad that such people say that science should be driven by data and results, but at the same time refuse to look at the actual results' (Bjørkeng 2015).

**12 The Bats Come Home to Roost in Silicon Valley**

My bets were settled in mid-2019. Our three judges, all physicists, agreed with my opponents that neither Brillouin nor Rossi had demonstrated evidence of LENR above 50% probability. They were more open than some of my opponents to the suggestion that the field met the lower bar, which recommended it for serious attention, given what was potentially at stake.

The Google-funded work had just been announced when the bets were finalised. I was delighted at this further evidence of interest in LENR in Silicon Valley. My opponents could take comfort from the fact that the Google team had failed to find excess heat, but from my point of view the more important thing was that they thought it worthwhile looking for it. As we saw in §8, the Google team were both well aware of, and strongly motivated by, the prudential argument for investigating LENR – the same argument that my critics had claimed to find unconvincing. Yet this seemed to cut no ice with those of my opponents who felt that LENR failed to reach the lower bar; by their lights, the Google team had been wasting their time.

My sense that the bats are coming home to roost for cold fusion sceptics in Silicon Valley has now received another boost. ICCF24, the 2022 meeting of the annual International Conference on Cold Fusion, is to be held in the Mountain View Museum of Computing, just up the road from the Google campus. The conference is being organised by the Anthropocene Institute (Anthropocene Institute 2021), whose President is Carl Page, brother of the Google co-founder, Larry Page. Carl Page has been a vocal supporter of LENR, as well as of other new nuclear technologies, such as molten salt fission reactors, for several years (Page 2016, 2019a).

In a talk from 2019 (Page 2019a, 2019b), Page describes how he studied the claims of LENR for more than a year, initially sceptical, before agreeing to speak to Robert Godes of Brillouin Energy (in which he is now an investor). This is one of Page's slides from that talk:

> **LENR: Cautious view of a stigmatized field.**
>
> Scientists must test their intellectual honesty from time to time by looking at research with conclusions outside the consensus. Otherwise how do you know you are a scientist, and not an adherent to an ideology? Or just fashionable.
>
> Given my interest in energy, I was asked to meet with a "cold fusion" researcher. I said "No" until I had a chance to see why the field is so unpopular with many intelligent people. After a year of reading and talking to experts, I discerned a textbook example of an important and unexpected result that provoked every form of unscientific reaction, literally terrorizing honest researchers. Motivated reasoning, rampant academic nepotism, self interest, intra-disciplinary conflict, ideology, math dominance, and authoritarian rule.

I have not discussed most of the 'unscientific' factors that Page has in mind here, but it would be naive to assume that they will not be relevant to the challenges of studying the potential risks of new technologies. The field needs a handbook for combatting those sorts of factors, too.

My recommendations in §11 had a less ambitious aim: improving the culture of discussion within science, so that it does a better job of studying low-probability high-impact risks. I have offered four recommendations for the science of extreme technological risk. All of them would also be beneficial elsewhere in science, in my view.

## 12 Final remarks on LENR

Some final remarks about the LENR case. I want to emphasise that the recent growth in interest in the field does not guarantee that cold fusion is real, let alone that it will turn out to be useful. What it does do, in my view, is to go some small way to addressing past failings, of two kinds. The lesser failing is lack of serious funding for the field over the past thirty years. The greater failing, responsible in large part for the lesser one, is the reputation trap.

I hope I've managed to convince some readers that in a case such as this, the reputation trap is a pathology of the scientific process. In my original *Aeon* piece I suggested that it amounted to shooting ourselves in the foot, but this doesn't quite capture what makes it pathological. A better image would be nailing our own feet to the floor – that combines unnecessary self-harm with self-imposed impediment to the exploration of an important search-space. Cold fusion will live on in the history of science as a classic example of how to get this wrong, in my view, even if it never boils a cup of tea.[7]

**Acknowledgements**

Many people have helped me with comments and discussion of this material at various points over the past ten years – so many, over such a long period, that I am sure that the following list is incomplete. With apologies to those who find themselves omitted, I'm grateful to everyone, named here or not, for their interest, comments and patience: Anthony Aguirre, Shahar Avin, María Baghramian, David Bailey, Philip Ball, Jonathan Borwein, Andrew Briggs, Hasok Chang, Clive Cookson, Adrian Currie, Finnur Dellsén, Heather Douglas, Eric Drexler, Luke Drury, Russ George, Masami Hayashi, Hugh Hunt, Yasuhiro Iwamura, Brian Josephson, Jirohta Kasagi, Charlie Kennel, Adrian Kent, Ross Koningstein, Tim Lewens, Michael McKubre, David Nagel, Seán Ó hÉigeartaigh, Ellen Quigley, Abbas Raza, Martin Rees, Carlo Rovelli, Carl Schulman, Mikey Slezak, Alan Smith, Jaan Tallinn, Francis Tanzella, Max Tegmark, Joris van der Schot, Alyssa Vance and Ken Wharton. I am also grateful to the editors of *Aeon* and *3QuarksDaily* for their willingness to publish my work on this topic, and to Clean Planet for their kind permission to reproduce Figure 1. Some of this work was made possible through the support of a grant on *Managing Extreme Technological Risk* from the Templeton World Charity Foundation. The opinions expressed here are my own and do not necessarily reflect the views of TWCF.**References**

Angner, E. (2020) 'Epistemic humility—knowing your limits in a pandemic', *Behavioural Scientist,* 13 April 2020. Available at: https://behavioralscientist.org/epistemic-humility-coronavirus-knowing-your-limits-in-a-pandemic/ (Accessed: 17 December 2021)

Anonymous (2016) 'Cold fusion is better left out in the cold', Editorial in *New Scientist,* 14 September 2016. Available at: https://www.newscientist.com/article/mg23130912-300-cold-fusion-is-better-left-out-in-the-cold/ (Accessed: 17 December 2021)

Anonymous (2019a) 'A Google programme failed to detect cold fusion — but is still a success', Editorial in *Nature,* 569, 599-600. Available at: https://doi.org/10.1038/d41586-019-01675-9 (Accessed: 17 December 2021)

Anonymous (2019b) 'Coming in from the cold', Editorial in *Nature Materials,* 18, 1145. Available at: https://www.nature.com/articles/s41563-019-0530-1 (Accessed: 27 January 2022)

Anthropocene Institute (2021) *About the Anthropocene Institute.* Available at: https://anthropoceneinstitute.com/about/ (Accessed: 17 December 2021)---

[7] The cup of tea reference is to a line from one of the field's well-known sceptics, the physicist Robert Park (Park 2004). McKubre replies that the field has actually boiled the equivalent of many thousands of cups of tea (Maguire 2014). He also recounts an incident in which Park refused to look at a paper offered by someone in the field, dropping it to the floor.

ARPA-E (2021a) *Who We Are.* Available at: https://arpa-e.energy.gov/about (Accessed: 17 December 2021)

ARPA-E (2021b) *Low-Energy Nuclear Reactions Workshop.* Available at: https://arpa-e.energy.gov/events/low-energy-nuclear-reactions-workshop (Accessed: 17 December 2021)

Bailey, D. and Borwein, J. (2014) 'Fusion energy: hope or hype?', Blogpost, 23 October 2014. Available at: https://www.huffpost.com/entry/fusion-energy-hope-or-hype_b_6031968 (Accessed: 17 December 2021)

Bailey, D. and Borwein, J. (2015) 'Cold fusion heats up: fusion energy and LENR update', Blogpost, 28 August 2015. Available at: https://www.huffpost.com/entry/post_10010_b_8052326 (Accessed: 17 December 2021)

Baldwin, M. (2015) *Making Nature: the History of a Scientific Journal,* Chicago: University of Chicago Press.

Ball, P. (2019a) 'Lessons from cold fusion, 30 years on', *Nature* 569, 601. Available at: https://www.nature.com/articles/d41586-019-01673-x (Accessed: 17 December 2021)

Ball, P. (2019b) 'Materials advances result from study of cold fusion*', MRS Bulletin* 44, 833–836. doi:10.1557/mrs.2019.260.

Bay Area Rationalists (2021) *Welcome to the Bay Area Rationalist Community.* Available at: http://www.bayrationality.com/ (Accessed: 17 December 2021)

Beiting, E. (2017) *Investigation of the nickel-hydrogen anomalous heat effect*, The Aerospace Corporation: Report No. ATR-2017-01760. Available at: https://www.lenr-canr.org/acrobat/BeitingEinvestigat.pdf (Accessed: 17 December 2021)

Berkes, H. (2012) 'Remembering Roger Boisjoly: he tried to stop shuttle Challenger launch', *NPR,* February 6, 2012. Available at:https://www.npr.org/sections/thetwo-way/2012/02/06/146490064/remembering-roger-boisjoly-he-tried-to-stop-shuttle-challenger-launch (Accessed: 17 December 2021)

Berlinguette, C.P., Chiang, YM., Munday, J.N. *et al* (2019) 'Revisiting the cold case of cold fusion', *Nature* 570, 45–51. Available at: https://doi.org/10.1038/s41586-019-1256-6 (Accessed: 17 December 2021)

Bjørkeng, P. K. (2015) '1 glass vann = energi til Hamar i et helt år?', *Aftenposten,* 12 September 2015. Available at: https://www.aftenposten.no/norge/i/4v8q/1-glass-vann-energi-til-hamar-i-et-helt-aar (Accessed: 17 December 2021)

Brooks, M. (2016) 'Cold fusion: science's most controversial technology is back', *New Scientist*, 14 September 2016. Available at: https://www.newscientist.com/article/mg23130910-300-cold-fusion-sciences-most-controversial-technology-is-back/ (Accessed: 17 December 2021)

Bushnell, D., Moses, R. Choi, S. (2021) 'Frontiers of space power and energy', NASA/TM–20210016143. Available at: https://ntrs.nasa.gov/api/citations/20210016143/downloads/NASA-TM-20210016143final.pdf (Accessed: 17 December 2021)

Clean Planet (2021) *Quantum Hydrogen Energy – Preparing for Commercial Application.* Available at: https://www.cleanplanet.co.jp/en/science/ (Accessed: 17 December 2021)

Cookson, C. (2012), 'Cold fusion: a personal history', *Financial Times,* 18 August 2012. Available at: https://www.ft.com/content/4f1c41e8-e66e-11e1-ac5f-00144feab49a (Accessed: 17 December 2021)

CORDIS (2021a) *Clean Energy from Hydrogen-Metal Systems.* Available at: https://cordis.europa.eu/project/id/951974 (Accessed: 17 December 2021)

CORDIS (2021b) *Breakthrough zero-emissions heat generation with hydrogen-metal systems.* Available at: https://cordis.europa.eu/project/id/952184 (Accessed: 17 December 2021)

CSER (2017) *Risk & the Culture of Science (Invite only workshop).* Available at: https://www.cser.ac.uk/events/risk-the-culture-of-science-invitation-only/ (Accessed: 17 December 2021)

DOE (2004) *Report of the Review of Low Energy Nuclear Reactions.* Available at: https://www.lenr-canr.org/acrobat/DOEreportofth.pdf (Accessed: 23 December 2021)

DOE (2021) *DOE Establishes New Office of Clean Energy Demonstrations Under the Bipartisan Infrastructure Law.* Available at: https://www.energy.gov/articles/doe-establishes-new-office-clean-energy-demonstrations-under-bipartisan-infrastructure-law (Accessed: 23 December 2021)

Douglas, H. (2000) 'Inductive risk and values in science', *Philosophy of Science,* 67(4), 559–579. https://www.journals.uchicago.edu/doi/abs/10.1086/392855

Douglas, H. (2007) 'Rejecting the Ideal of Value-Free Science', in Kincaid, K., Dupré, J. and Wylie, A. (eds.), *Value-Free Science? Ideals and Illusions*, Oxford: Oxford University Press, 120–141.

Douglas, H. (2009) *Science, Policy, and the Value-Free Ideal.* Pittsburgh: University of Pittsburgh Press.

Dumaine, B. (2015) 'This investor is chasing a new kind of fusion', *Fortune,* 28 September 2015. Available at: https://fortune.com/2015/09/27/ceo-cherokee-investment-partners-low-energy-nuclear-reaction/ (Accessed: 17 December 2021)

ENEA (2013) 'New advancements on the Fleischmann-Pons Effect: paving the way for a potential new clean renewable energy source?', Workshop at the European Parliament, Brussels, 3 June 2013. Available at: https://www.enea.it/it/Ufficio-Bruxelles/news/new-advancements-on-the-fleischmann-pons-effect-paving-the-way-for-a-potential-new-clean-renewable-energy-source (Accessed: 17 December 2021)

Fleischmann, M. and Pons, S. (1989) 'Electrochemically induced nuclear fusion of deuterium', *Journal of Electroanalytical Chemistry and Interfacial Electrochemistry*, 261, 301–308. doi.org/10.1016/0022-0728(89)80006-3

Goodstein, D. (1994) 'Pariah science: whatever happened to cold fusion? *The American Scholar*, 63:4, 527–541. Available at: https://www.its.caltech.edu/~dg/fusion_art.html (Accessed: 27 January 2022)

Greco, K. (2021) 'Review of 1989 and 2004 DOE Reports'. Available at: https://arpa-e.energy.gov/sites/default/files/2021LENR_workshop_Greco.pdf (Accessed: 17 December 2021)

Habgood-Coote, J. (2018) 'Stop talking about fake news!', *Medium,* 16 July 2018. Available at: https://medium.com/@josh_coote/stop-talking-about-fake-news-cacf90998566 (Accessed: 17 December 2021)

Habgood-Coote (2019 'Stop talking about fake news!', *Inquiry: an Interdisciplinary Journal of Philosophy,* 62, 1033–1065.

Hambling, D. (2011) 'Success for Andrea Rossi's E-Cat cold fusion system, but mysteries remain', *Wired,* 29 October 2011. Available at: https://www.wired.co.uk/article/rossi-success (Accessed: 17 December 2021)

Hasvold, J. (2018) 'Condensed Matter Nuclear Science – fiksjon eller virkelighet (fiction or reality)', FFI-RAPPORT 18/00678. Available at: https://publications.ffi.no/nb/item/asset/dspace:4203/18-00678.pdf (Accessed: 17 December 2021)

Hughes, N. (2020) 'NASA Detects Lattice Confinement Fusion', *YouTube* video, 1 August 2020. Available at: https://www.youtube.com/watch?v=e2pcrgFb7L4 (Accessed: 17 December 2021)

ICCF-24 (2021) *ICCF-24 Silicon Valley*. Available at: https://www.iccf24.org/ (Accessed: 17 December 2021)

Jones, S., Palmer, E., Czirr, J., *et al* (1989) 'Observation of cold nuclear fusion in condensed matter', *Nature,* 338, 737–740. https://doi.org/10.1038/338737a0

Kaneko, K. (2016) 'Successful reproduction of patents in the US accelerates re-evaluation of "cold fusion"', *Nikkei,* 9 September 2016. Available at: https://www.nikkei.com/article/DGXMZO06252800Z10C16A8000000/ (Accessed: 23 December 2021 via Google Translate)

Kasagi, J. and Iwamura, Y. (2008) 'Country history of Japanese work on cold fusion', in *ICCF-14 International Conference on Condensed Matter Nuclear Science,* Washington, DC. Available at: https://www.lenr-canr.org/acrobat/KasagiJcountryhis.pdf (Accessed: 11 January 2022)

Kitamura, A., Takahashi, A., Takahashi, K., Seto, R., Hatano, T., *et al* (2018) 'Excess heat evolution from nanocomposite samples under exposure to hydrogen isotope gases', *International Journal of Hydrogen Energy,* 43(33), 16187–16200. https://doi.org/10.1016/j.ijhydene.2018.06.187.

Koningstein, R. and Fork, D. (2014) 'What it would really take to reverse climate change', *IEEE Spectrum,* 18 November 2014. Available at: https://spectrum.ieee.org/what-it-would-really-take-to-reverse-climate-change (Accessed: 17 December 2021)

Koziol, M. (2020) 'Spacecraft of the future could be powered by lattice confinement fusion – NASA researchers demonstrate the ability to fuse atoms inside room-temperature metals', *IEEE Spectrum,* 5 August 2020. Available at: https://spectrum.ieee.org/nuclear-fusiontokamak-not-included (Accessed: 17 December 2021)

Koziol, M. (2021) 'Whether cold fusion or low-energy nuclear reactions, U.S. Navy researchers reopen case', *IEEE Spectrum,* 22 March 2021. Available at: https://spectrum.ieee.org/cold-fusion-or-low-energy-nuclear-reactions-us-navy-researchers-reopen-case (Accessed: 17 December 2021)

Kuhn, T. (1962) *The structure of scientific revolutions.* Chicago: University of Chicago Press.

Levi, G., Foschi, E., Hartman, T., Höistad, B., Petterson, R., Tegner, Essen, H. (2013) 'Indication of anomalous heat energy production in a reactor device', Available at: https://arxiv.org/abs/1305.3913v (Accessed: 17 December 2021)

Levi, G., Foschi, E., Hartman, T., Höistad, B., Petterson, R., Tegner, Essen, H. (2014) 'Observation of abundant heat production from a reactor device and of isotopic changes in the fuel', Preprint. Available at: http://www.sifferkoll.se/sifferkoll/wp-content/uploads/2014/10/LuganoReportSubmit.pdf (Accessed: 17 December 2021)

LENR-CANR.ORG (2021) *LENR-CANR.ORG – A library of papers about cold fusion.* Available at: https://lenr-canr.org/ (Accessed: 17 December 2021)

Lewan, M. (2015a) 'Replication attempts are heating up cold fusion', Blogpost, 1 February 2015. Available at: https://animpossibleinvention.com/2015/02/01/replication-attempts-are-heating-up-cold-fusion/ (Accessed: 17 December 2021)

Lewan, M. (2015b) 'Rossi has been granted US patent on the E-Cat — fuel mix specified', Blogpost, 25 August 2015. Available at: https://animpossibleinvention.com/2015/02/01/replication-attempts-are-heating-up-cold-fusion/ (Accessed: 17 December 2021)

Lewan, M. (2015c) 'Swedish scientists claim LENR explanation break-through', Blogpost, 15 October 2015. Available at: https://animpossibleinvention.com/2015/10/15/swedish-scientists-claim-lenr-explanation-break-through/ (Accessed: 17 December 2021)

Lucretius (1916) *De rerum natura*. Translated by William Ellery Leonard, New York: E. P. Dutton.

Lundin, R. and Lidgren, H. (2015) 'Nuclear spallation and neutron capture induced by ponderomotive wave forcing. *IRF scientific report*, vol. 305. ISSN 0284-1703. Available at: https://www.semanticscholar.org/paper/Nuclear-Spallation-and-Neutron-Capture-Induced-by-Lundin-Lidgren/19d09a644d8190ba899b0e7588341a4787c27246 (Accessed: 17 December 2021)

Maguire, J. (2014) 'Dr. Michael McKubre: experimental cold fusion, pseudo-skepticism, and progressing LENR', *Q-Niverse Podcast,* 25 April 2014. Available at: https://www.podomatic.com/podcasts/jmag0904/episodes/2014-04-24T11_40_44-07_00 (Accessed: 17 December 2021)

Marwan, J., McKubre, M., Tanzella, F., Hagelstein, P., Miles, M, Schwarz, M. et al (2010) 'A new look at low-energy nuclear reaction (LENR) research: a response to Shanahan', *Journal of Environmental Monitoring,* 12, 1765–1770. http://dx.doi.org/10.1039/c0em00267d

Malle, L. (1981) *My Dinner with Andre*. Movie, Saga Productions.

McDonough, W. (2015) 'Investing in innovation for the common good', Webpage, Available at: https://mcdonough.com/writings/investing-innovation-common-good/ (Accessed: 17 December 2021)

McKay, P. (2017) *Disagreement in Science and Beyond – A workshop organised by WEXD (Dublin) and CSER (Cambridge).* Available at: http://whenexpertsdisagree.ucd.ie/disagreement_in_science_and_beyond/ (Accessed: 17 December 2021)

McKubre, M. (2009) 'Cold fusion (LENR) – One perspective on the state of the science", *15th International Conference on Condensed Matter Nuclear Science,* Rome, Italy: ENEA. Available at: https://www.lenr-canr.org/acrobat/McKubreMCHcoldfusionb.pdf (Accessed: 17 December 2021)

McKubre, M. (2019) 'Critique of *Nature* Perspective Article on Google-sponsored Pd-D and Ni-H experiments', *Infinite Energy,* Issue 146, July/August 2019, 1-4. Available at: https://www.infinite-energy.com/iemagazine/issue146/McKubreGoogle.pdf (Accessed: 17 December 2021)

Metzler, T. (2021) 'Towards a LENR reference experiment', Workshop presentation, 21 October 2021. Available at: https://www.youtube.com/watch?v=Ec9OnfWvOjs&t=1726s (Accessed: 29 December 2021)

Mizuno, T. (2017) 'Observation of excess heat by activated metal and deuterium gas', *J. Condensed Matter Nucl. Sci.,* 25, 1–25. https://www.lenr-canr.org/acrobat/MizunoTpreprintob.pdf

Miura (2021) *MIURA CO. LTD. and Clean Planet Inc. conclude an agreement for joint development of industrial boilers that use Quantum Hydrogen Energy,* 28 September 2021. Available at: https://www.miuraz.co.jp/news/newsrelease/2021/1132.php (Accessed: 17 December 2021)

NASA Glenn Research Center (2021) 'Lattice confinement fusion', 15 April 2020. Available at: https://www1.grc.nasa.gov/space/science/lattice-confinement-fusion/ (Accessed: 17 December 2021)

NikkeiBP (2021) 'Boilers using "nuclear fusion /heat" are put into practical use, and heat is taken out with metal laminated chips. Jointly developed by Miura Co., Ltd. and Clean Planet, commercialized in 2023', 4 October 2021. Available at: https://project.nikkeibp.co.jp/ms/atcl/19/news/00001/02043/?ST=msb&P=2 (Accessed: 17 December 2021 via Google Translate)

NYU Philosophy (2021) *NYU Guidelines for Respectful Philosophical Discussion.* Available at: https://as.nyu.edu/content/nyu-as/as/departments/philosophy/climate/initiatives/nyu-guidelines-for-respectful-philosophical-discussion.html (Accessed: 17 December 2021)

Page, C. (2016) 'Low Energy Nuclear Reactions Work and Could Supplant Fossil Fuels', *Edge* 2016. Available at: https://www.edge.org/response-detail/26753?fbclid=IwAR3iKDxvjbnZlo1FEcyrnvbQK6F4AHUiXfCfhTO_CucuxosR-kn03n9CKOY (Accessed: 17 December 2021)

Page, C. (2019a) 'Context and thoughts on LANR/LENR', Slide presentation, 24 March 2019. Available at: https://www.lenr-canr.org/acrobat/PageCcontexttho.pdf (Accessed: 17 December 2021)

Page, C. (2019b) 'Context and thoughts on LANR/LENR', YouTube, 16 May 2019. Available at: https://www.youtube.com/watch?v=Oebkw0G7tEw (Accessed: 17 December 2021)

Park, R. (2004). 'Cold fusion: just when you think life can't get any sillier', *What's New,* 2 April 2004. Available at: http://bobpark.physics.umd.edu/WN04/wn080604.html (Accessed: 17 December 2021

Pines, V., Pines, M., Chait, A., Steinetz, B. M., Forsley, L. P., Hendricks, R. C. et al (2020) 'Nuclear fusion reactions in deuterated metals', *Phys. Rev. C,* 101(4), 044609. DOI: 10.1103/PhysRevC.101.044609.

Price, H. (2015) 'Why do scientists dismiss the possibility of cold fusion?', *Aeon,* 21 December 2015. Available at: https://aeon.co/essays/why-do-scientists-dismiss-the-possibility-of-cold-fusion (Accessed: 17 December 2021)

Price, H. (2016) 'Is the cold fusion egg about to hatch?', *Aeon,* 24 March 2016. Available at: https://aeon.co/ideas/is-the-cold-fusion-egg-about-to-hatch (Accessed: 17 December 2021)

Price, H. (2019) 'Icebergs In the room? Cold fusion at 30', *3 Quarks Daily,* 4 March 2019. Available at: https://3quarksdaily.com/3quarksdaily/2019/03/icebergs-in-the-room-cold-fusion-at-thirty.html (Accessed: 17 December 2021)

Reich, E. (2011) 'Speedy neutrinos challenge physicists', *Nature News*, 477(7366): 520. doi:10.1038/477520a

Reich, E. (2012) 'Flaws found in faster-than-light neutrino measurement', *Nature News*, doi:10.1038/nature.2012.10099.

Rhodes, R. (1986) *The Making of the Atomic Bomb*, New York: Simon and Schuster.

Russell, E. (2018) 'Francis Tanzella on the Cold Fusion Now! Podcast', *Cold Fusion Now!,* 13 October 2018. Available at: https://coldfusionnow.org/francis-tanzella-on-the-cold-fusion-now-podcast/ (Accessed: 17 December 2021)

Salamon, M., Wrenn, M., Bergeson, H. *et al* (1990) 'Limits on the emission of neutrons, γ-rays, electrons and protons from Pons/Fleischmann electrolytic cells', *Nature* 344, 401–405. doi.org/10.1038/344401a0

Scott, R. (2015) *The Martian.* Movie, Scott Free Productions.

Steinetz, B., Benyo, T., Chait, A., Hendricks, R., Forsley, L., Baramsai, B., *et al* (2020) 'Novel nuclear reactions observed in bremsstrahlung-irradiated deuterated metals', *Phys. Rev. C,* 101(4), 044610. DOI: 10.1103/PhysRevC.101.044610.

Szpak, S., *et al* (2008) 'SPAWAR Systems Center-Pacific Pd:D Co-Deposition Research: Overview of Refereed LENR Publications', in *ICCF-14 International Conference on Condensed Matter Nuclear Science,* Washington, DC. Available at: https://www.researchgate.net/publication/242327687_SPAWAR_Systems_Center-Pacific_PdD_CoDeposition_Research_Overview_of_Refereed_LENR_Publications (Accessed: 17 December 2021)

Takahashi, A., Kitamura, A., Takahashi, K., Seto, R., Matsuda, Y., Iwamura, Y., *et al* (2017) 'Phenomenology and controllability of new exothermic reaction between metal and hydrogen', *Brief summary report of MHE Project Japan for 2015 October – 2017 October.* Available at: https://www.researchgate.net/publication/322160963_Brief_Summary_Report_of_MHE_Project_Phenomenology_and_Controllability_of_New_Exothermic_Reaction_between_Metal_and_Hydrogen (Accessed: 17 December 2021)

Tanzella, F. (2016) *Isoperibolic Hydrogen Hot Tube Reactor Studies: Interim progress report for the period 1 March – 5 December 2016,* SRI International Project P21429. https://brillouinenergy.com/newwebsite/wp-content/uploads/2018/12/SRI_ProgressReport.pdf (Accessed: 17 December 2021)

Tanzella, F. (2018a) *Isoperibolic Hydrogen Hot Tube Reactor Studies: Technical progress report for the period 1 January – 31 December 2017,* SRI International Project P21429. Available at: https://brillouinenergy.com/newwebsite/wp-content/uploads/2018/12/SRI_Technical_Report.pdf (Accessed: 17 December 2021)

Tanzella, F. (2018b) *Isoperibolic Hydrogen Hot Tube Reactor Studies: Final progress report for the July 1st 2016 through December 31st 2018,* SRI International Project P21429. Available at: https://brillouinenergy.com/newwebsite/wp-content/uploads/2019/04/Brillouin-SRI-Technical-Progress-Report-Final-Public-2018.pdf (Accessed: 17 December 2021)

UBC (2021) *Curtis Berlinguette.* Available at: https://groups.chem.ubc.ca/cberling/curtis-berlinguette/ (Accessed: 17 December 2021)

Vineberg, S. (2016) 'Dutch Book Arguments', in Zalta, E. (ed.), *The Stanford Encyclopedia of Philosophy* (Spring 2016 Edition). Available at: https://plato.stanford.edu/archives/spr2016/entries/dutch-book/ (Accessed: 17 December 2021)

Wang, B. (2015) 'China's LENR is getting excess 600 watts of heat from 780 watts of input power', Blogpost, 8 June 2015. Available at: https://www.nextbigfuture.com/2015/06/chinas-lenr-is-getting-excess-600-watts.html (Accessed: 17 December 2021)

Wells, D., Campbell, R., Chase, A., Daniel, J., Darling, M., Green, C., *et al* (2014) 'Low Energy Nuclear Reaction Aircraft— 2013 ARMD Seedling Fund Phase I Project', NASA/TM–2014-218283. Available at: https://nari.arc.nasa.gov/sites/default/files/Wells_TM2014-218283%20Low%20Energy%20Nuclear%20Reaction%20Aircraft_0.pdf (Accessed: 17 December 2021)

Wikipedia contributors (2021a) 'Sergio Focardi', in *Wikipedia, The Free Encyclopedia.* Retrieved 23:54, December 22, 2021, from https://en.wikipedia.org/w/index.php?title=Sergio_Focardi&oldid=997654571

Wikipedia contributors (2021b) 'Cold fusion', in *Wikipedia, The Free Encyclopedia*. Retrieved 23:29, December 10, 2021, from
https://en.wikipedia.org/w/index.php?title=Cold_fusion&oldid=1059561939

Wikipedia contributors (2021c) 'Space Shuttle Challenger disaster', in *Wikipedia, The Free Encyclopedia*. Retrieved 04:06, December 11, 2021, from
https://en.wikipedia.org/w/index.php?title=Space_Shuttle_Challenger_disaster&oldid=1059690149

Wikipedia contributors (2021d) 'Ad hominem', in *Wikipedia, The Free Encyclopedia*. Retrieved 04:44, December 30, 2021, from
https://en.wikipedia.org/w/index.php?title=Ad_hominem&oldid=1061366737

Zeldovich, L. (2019) 'A history of human waste as fertilizer', *JSTOR Daily,* 18 November 2019. Available at: https://daily.jstor.org/a-history-of-human-waste-as-fertilizer/ (Accessed: 17 December 2021)